\documentclass[epic,eepic,authoryear]{elsarticle}
\usepackage{epsfig,natbib,lineno,amsfonts}
\usepackage{multirow}
\usepackage{setspace}
\usepackage{amsmath,amssymb}
\usepackage{bm}
\usepackage{multirow} 
\usepackage{color}
\usepackage{fullpage}
\usepackage[english]{babel}
\usepackage{placeins}

\newcommand*\patchAmsMathEnvironmentForLineno[1]{
 \expandafter\let\csname old#1\expandafter\endcsname\csname #1\endcsname
 \expandafter\let\csname oldend#1\expandafter\endcsname\csname end#1\endcsname
 \renewenvironment{#1}
 {\linenomath\csname old#1\endcsname}
 {\csname oldend#1\endcsname\endlinenomath}}
\newcommand*\patchBothAmsMathEnvironmentsForLineno[1]{
 \patchAmsMathEnvironmentForLineno{#1}
 \patchAmsMathEnvironmentForLineno{#1*}}
\AtBeginDocument{
\patchBothAmsMathEnvironmentsForLineno{equation}
\patchBothAmsMathEnvironmentsForLineno{align}}

\begin{document}
%%\begin{linenumbers}
\begin{frontmatter}
\title{Modeling cross-hole slug tests in an unconfined aquifer}

\author[cp]{Bwalya Malama\corref{cor}}
\ead{bmalama@scalpoly.edu}
\author[snl]{Kristopher L. Kuhlman}
\author[afc]{Ralf Brauchler}
\author[eth]{Peter Bayer}
\cortext[cor]{Corresponding author. Tel.: + 1 805 756 2971; fax: + 1 805 756 1402}
\address[cp]{Natural Resources Management \& Environmental Sciences Department,\\ California Polytechnic State University, San Luis Obispo, California}
\address[snl]{Sandia National Laboratories, Albuquerque, New Mexico}
\address[afc]{AF-Consult Switzerland Ltd, Baden, Switzerland}
\address[eth]{ETH Z{\"u}rich, Z{\"u}rich, Switzerland.}

\begin{abstract}
A modified version of a published slug test model for unconfined aquifers is applied to cross-hole slug test data collected in field tests conducted at the Widen site in Switzerland. The model accounts for water-table effects using the linearised kinematic condition. The model also accounts for inertial effects in source and observation wells. The primary objective of this work is to demonstrate applicability of this semi-analytical model to multi-well and multi-level pneumatic slug tests. The pneumatic perturbation was applied at discrete intervals in a source well and monitored at discrete vertical intervals in observation wells. The source and observation well pairs were separated by distances of up to 4~m. The analysis yielded vertical profiles of hydraulic conductivity, specific storage, and specific yield at observation well locations. The hydraulic parameter estimates are compared to results from prior pumping and single-well slug tests conducted at the site, as well as to estimates from particle size analyses of sediment collected from boreholes during well installation. The results are in general agreement with results from prior tests and are indicative of a sand and gravel aquifer. Sensitivity analysis show that model identification of specific yield is strongest at late-time. However, the usefulness of late-time data is limited due to the low signal-to-noise ratios.
\end{abstract}

\end{frontmatter}

\section{Introduction}
Slug tests are a common tool in hydrogeology for hydraulic characterization of aquifers because they are quick, obviate the need for waste water disposal, require less equipment, and are not as labor intensive as pumping tests. Fundamentally, they involve instantaneous (step) perturbation of fluid pressure in an interval followed by continuous monitoring of the pressure change as it dissipates by fluid flow through the aquifer. This is typically achieved by either dropping a slug mass into a well \citep{cooper1967} or pneumatically pressurizing the water column in a well \citep{butler1998, malama2011}, a configuration referred to as a single well test. Several mathematical models are available in the hydrogeology literature for analyzing confined \citep{cooper1967, bredehoeft1980, zurbuchen2002, butler2004} and unconfined \citep{bouwer1976, springer1991, hyder1994, spane1996a, zlotnik1998, malama2011} aquifer slug test data under the Darcian flow regime. Consideration of slug tests under non-Darcian flow regimes may be found in \citet{quinn2013} and \citet{wang2015}.

Slug tests have the advantage of only involving limited contact with and minimal disposal of effluent formation water. As such, they have found wide application for characterizing heterogeneous formations at contaminated sites \citep{shapiro1998} and for investigating flow in fractured rock \citep{quinn2013, ji2015, ostendorf2015}. However, the small volumes of water involved impose a physical limit on the volume of the formation interrogated during tests \citep{shapiro1998, beckie2002} because the resulting pressure perturbations often do not propagate far enough to be measurable in observation wells. As a result, hydraulic parameters estimated from single well slug-test data can only be associated with the formation volume within the immediate vicinity of the source well \citep{beckie2002, butler2005}.

Cross-hole (or multi-well) slug tests are less common but have been applied to interrogate relatively large formation volumes in what has come to be known as hydraulic tomography \citep{yeh2000, illman2009}. For example, \citet{vesselinov2001b} and \citet{illman2001} used pneumatic cross-hole injection tests to hydraulically characterized a fractured unsaturated rock formation with dimensions of $30\times30\times30$~$\mathrm{m}^3$. \citet{barker1983} presented evidence of measurable pressure responses in observation wells several meters from the source well. \citet{audouin2008} reported cross-hole slug tests conducted in fractured rock, where they collected data in observations wells at radial distances 30 to about 120~m from the source well, and observed maximum peak amplitudes ranging from 5 to 20~cm. This demonstrated empirically that slug test pressure perturbations can propagate over relatively large distances beyond the immediate vicinity of the source well, albeit for fractured rocks, which have high hydraulic diffusivities. \citet{brauchler2010} attempted to intensively apply cross-hole slug tests to obtained a detailed image of confined aquifer heterogeneity. They used the model of \citet{butler2004} to estimate aquifer hydraulic conductivity, specific storage and anisotropy. Cross-hole slug tests in unconfined aquifers, neglecting wellbore inertial effects, have been reported by \citet{spane1996a}, \citet{spane1996b}, and \citet{belitz1999} for source-to-observation well distances not exceeding 15 m.

Recently \citet{paradis2013} and \citet{paradis2014, paradis2015} analysed synthetic cross-hole slug test data using a model for over-damped observation well responses. The need, therefore, still exists to analyse field data and characterize high permeability heterogeneous unconfined aquifers using cross-hole slug tests where source and observation well inertial effects may not be neglected. \citet{malama2011} developed a slug test model for unconfined aquifers using the linearised kinematic condition of \citet{neuman1972} at the water-table, and accounting for inertial effects of the source well. They analysed data from single-well tests performed in a shallow unconfined aquifer. This work extends the application of the model of \citet{malama2011} to multi-well tests and to response data collected in observation wells. The data analysed were collected at multiple vertical intervals in an observation well about 4 m from the source well, which itself was perturbed at multiple intervals. The model and data are used to estimate hydraulic conductivity, specific storage, and specific yield. The sensitivity of predicted model behaviour to these parameters is also analysed. In the following, the mathematical model is presented, the multi-level multi-well tests are described, and data analysed. The work concludes with an analysis of the sensitivity coefficients for the hydraulic and well characteristic parameters.

\begin{figure}[h] % Figure 1
\includegraphics[width=0.75\textwidth]{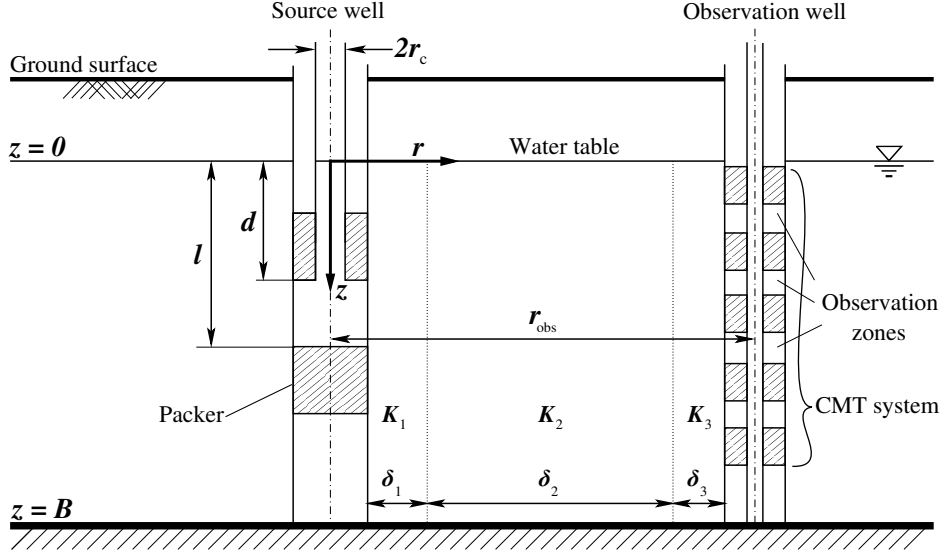} 
\caption{\label{fig:slugschematic} Schematic of a typical cross-hole slug test set-up for an unconfined aquifer. For the tests reported herein, the source and observation intervals were isolated with a multi chamber well not a multi-packer system.}
\end{figure}

\section{Slug Test Model}
\citet{malama2011} developed a model for formation and source well response to slug tests performed in unconfined aquifers using the linearised kinematic condition at the water-table. The model allows for estimation of specific yield in addition to hydraulic conductivity and specific storage. The model also accounts for source-well wellbore storage and inertial effects. Wellbore storage in the source well is treated in the manner of \citet{cooper1967}. A schematic of the conceptual model used to derive the semi-analytical solution is shown in Figure \ref{fig:slugschematic}. Whereas the solution of \citep{malama2011} was obtained for and applied to source wells, here a more complete solution is presented that applies to observation wells. The complete aquifer response for both source and observation wells is given by (see Appendix~A and \citet{malama2011} for details)
\begin{equation}
\label{eqn:}
\hat{\overline{s}}_D = \hat{\overline{u}}_D\left\{
  \begin{split}
   \left[1 - \hat{\overline{v}}_D\left(d_D\right)\right] \hat{\overline{f}}_1(z_D) & \quad \mbox{$\forall z_D\in[0,d_D]$}\\
   1 - \hat{\overline{v}}_D & \quad \mbox{$\forall z_D\in[d_D,l_D]$}\\
   \left[1 - \hat{\overline{v}}_D\left(l_D\right)\right]\hat{\overline{f}}_2(z_D) & \quad \mbox{$\forall z_D\in[l_D,1]$},
  \end{split} \right.
\end{equation}
where $\hat{\overline{s}}_D$ is the double Laplace-Hankel transform of the dimensionless formation head response $s_D =s/H_0$, $d_D=d/B$ and $l_D=l/B$ are dimensionless depths to the top and bottom of the test interval, $z_D=z/B$ ($z\in [0,B]$) is dimensionless depth below the water-table, $B$ is initial saturated thickness,
\begin{equation}
\label{eqn:uD}
\hat{\overline{u}}_D = \frac{C_D(1-p\overline{H}_D)}{\kappa\eta^2\xi_w\mathrm{K}_1(\xi_w)},
\end{equation}
\begin{equation}
\label{eqn:vD}
 \hat{\overline{v}}_D = \frac{\Delta_0(d_D)}{\Delta_0(1)} \cosh(\eta z_D^\ast) + \sinh(\eta \l_D^\ast)\frac{\Delta_0' (z_D)}{\eta\Delta_0(1)},
\end{equation}
\begin{align}
\hat{\overline{f}}_1(z_D) & = \frac{\Delta_0'(z_D)}{\Delta_0'(d_D)} ,\\
\hat{\overline{f}}_2(z_D) & = \frac{\cosh(\eta z_D^\ast)}{\cosh(\eta l_D^\ast)} ,
\end{align}
\begin{equation}
\Delta_0 (z_D) = \sinh(\eta z_D) + \varepsilon \cosh(\eta z_D),
\end{equation}
and
\begin{equation}
\Delta'_0 (z_D) = \eta \left[\cosh\left(\eta z_D\right) + \varepsilon \sinh\left(\eta z_D\right) \right].
\end{equation}
Additionally, $z_D^\ast=1-z_D$, $l_D^\ast = 1 - l_D$, $\eta = \sqrt{(p+a_i^2)/\kappa}$, $p$ and $a_i$ are the dimensionless Laplace and finite Hankel transform parameters, $C_D=r_{D,c}^2/(b_s S_s)$ is the dimensionless wellbore storage coefficient of the source well, $S_s$ is formation specific (elastic) storage, $b_s=l-d$ the length of the source well completion interval, $\kappa = K_z/K_r$ is the formation anisotropy ratio, $K_z$ and $K_r$ are vertical and radial hydraulic conductivities, $\xi_w=r_{D,w}\sqrt{p}$, $\varepsilon=p/(\eta\alpha_D)$, and $\mathrm{K}_1()$ is the first-order second-kind modified Bessel function \citep[\S 10.25]{Olver:2010:NHMF}. The relevant dimensionless parameters are listed in Table \ref{tab:dimlessparameters}.

\begin{table}[ht] % Table 1
\caption{\label{tab:dimlessparameters}Dimensionless variables and parameters}
\centering
\begin{tabular}{lll}
\hline 
$s_{D,i}$ &=& $s_i/H_0$\\
$H_D$ &=& $H(t)/H_0$\\
$r_D$ &=& $r/B$\\
$r_{D,w}$ &=& $r_w/B$\\
$r_{D,c}$ &=& $r_c/B$\\
$r_{D,s}$ &=& $r_s/B$\\
$R_D$ &=& $R/B$\\
$z_D$ &=& $z/B$\\
$d_D$ &=& $d/B$\\
$t_D$ &=& $\alpha_{r,1} t/B^2$\\
$C_{D}$ &=& $r_{D,c}^2/(bS_s)$\\
$\alpha_{D}$ &=& $\kappa\sigma$\\
$\beta_1$ &=& $8\nu L/(r_c^2 g T_c)$\\
$\beta_2$ &=& $L_e/(g T_c^2)$\\
$\beta_D$ &=& $\beta_1/\sqrt{\beta_2}$\\
$\kappa_i$ &=& $K_{z,i}/K_{r,i}$\\
$\sigma$ &=& $BS_s/S_y$\\
$\gamma$ &=& $K_{r,2}/K_{r,1}$\\
$\vartheta$ &=& $2b S_{s,2} (r_w/r_c)^2$\\
$\xi_\mathrm{sk}$ &=& $r_\mathrm{sk}/r_w$\\
$\xi_w$ &=& $r_{D,w}\sqrt{p}$\\
$\eta^2$ &=& $(p+a_i^2)/\kappa$\\
\hline
\end{tabular} 
\end{table}

The function $\overline{H}_D(p)$ in \eqref{eqn:uD} is the Laplace transform of $H_D(t_D) = H(t)/H_0$, the normalized response in the source well, and is given by
\begin{align}
\label{eqn:solution2}
\overline{H}_D(p) & = \frac{\overline{\psi}_1(p)}{\omega_{D,s}^2 + p\overline{\psi}_1(p)},
\end{align}
where $\omega_{D,s} = \omega_s T_c$, $\omega_s = \sqrt{g/L_e}$ is the source well frequency, $L_e$ is a characteristic length associated with the source well oscillatory term, $T_c = B^2/\alpha_r$ is a characteristic time, $g$ is the acceleration due to gravity, and
\begin{align}
\overline{\psi}_1(p) & = p + \gamma_{D,s} + \frac{\omega_{D,s}^2}{2}\overline{\Omega}\left(r_{D,w},p\right).
\end{align}
The function $\overline{\Omega}$ is defined by
\begin{equation}
\label{eqn:Lap_Omega}
\overline{\Omega}(r_{D,w},p) = \left.\mathsf{H}_0^{-1}\left\{\hat{\overline{\Omega}} \left(a_i,p \right)\right\}\right|_{r_{D,w}},
\end{equation}
where $\mathsf{H}_0^{-1}\{\}$ denotes the inverse zeroth-order finite Hankel transform operator, $r_{D,w} = r_w/B$ is the dimensionless wellbore radius, $\gamma_{D,s} = \gamma_s T_c$, $\gamma_s$ is the source well damping coefficient, and
\begin{equation}
\label{eqn:HankLap_Omega}
\hat{\overline{\Omega}} (a_i,p) = \frac{C_D \left[1 - \left\langle \hat{\overline{w}}_D \left(a_i,p \right) \right \rangle \right]}{\kappa \eta^2 \xi_w \mathrm{K}_1(\xi_w)}.
\end{equation}
\citet{malama2011} showed that
\begin{equation}
\label{eqn:wD}
\langle \hat{\overline{w}}_D \rangle = \frac{1}{\eta b_{D,s}} \frac{\Delta_0(d_D)}{\Delta_0(1)} \left\{\sinh(\eta d_D^\ast) - \left[2 - \frac{\Delta_0(l_D)}{\Delta_0(d_D)}\right] \sinh(\eta l_D^\ast)\right\},
\end{equation}
where $d_D^\ast = 1-d_D$, and $b_{D,s}=b_s/B$. According to \citet{butler2004}, the source well damping coefficient is $\gamma_s = 8\nu L /(L_e r_c^2)$, where $\nu$ is the kinematic viscosity of water and $L$ is a characteristic length associated with the perturbed column of water in the source well.

Whereas \cite{malama2011} used the infinite Hankel transform, here a finite Hankel transform \citep{sneddon95,miles71} is used for inversion, with the transform pair defined as
\begin{align}
  \label{eq:finite-Hankel-pair}
  \hat{f}(a_i) &= \mathsf{H}_0 \left\{ f(r_D)\right\} = \int_0^{R_D} r_D f(r_D) \mathrm{J}_0 (r_D a_i) \; \mathrm{d}r_D, \nonumber\\
  f(r_D) &= \mathsf{H}^{-1}_0 \left\{ \hat{f}(a_i)\right\} = \frac{2}{R_D^2}\sum_{i=0}^{\infty} \hat{f}(a_i) \frac{ \mathrm{J}_0(r_D a_i)}{\left[ \mathrm{J}_1 (R_D a_i)\right]^2},
\end{align}
where $a_i$ are the roots of $\mathrm{J}_0(R_D a_i)=0$, $R_D = R/B$, $R$ is the radius of influence of the source well, and $\mathrm{J}_n()$ is the $n$th-order first-kind Bessel function \citep[\S 10.2]{Olver:2010:NHMF}. For the specified roots and Hankel transform pair given in \eqref{eq:finite-Hankel-pair}, a homogeneous Dirichlet boundary condition is enforced at $r_D=R_D$. Due to the short duration of the signal, a radius of influence such that $R \ge 2 r_\mathrm{obs}$ is sufficient. The finite Hankel transform is chosen for computational expedience; it is simpler to invert numerically than the infinite Hankel transform \citep{malama13}. Laplace transform inversion is performed using the algorithm of \cite{dehoog1982}. The software used to implement the analytical solution described here is released under an open-source MIT license and is available from a public Bitbucket repository (\texttt{https://bitbucket.org/klkuhlm/slug-osc}).

\subsection{Approximation of observation well skin}
It is assumed here that the slug test response at the observation well is due to fluid flow through the sub-domains associated with the source and observation wells and the formation shown in Figure~\ref{fig:slugschematic}. The well skin and formation hydraulic conductivities, $K_i$, $i=1,2,3$, are arranged in series for radial flow, and in parallel for vertical flow. Hence, the effective radial and vertical hydraulic conductivity, $\langle K_r \rangle$ and $\langle K_z \rangle$, of the formation between the source and observation wells are approximated as $$\langle K_r \rangle = \delta_T/\sum_{n = 1}^3 \frac{\delta_i^\ast}{K_i},$$ and $$\langle K_z \rangle = \frac{1}{\delta_T} \sum_{n=1}^3 \delta_i K_i,$$
where $\delta_1^\ast= (\hat{r}/r_1)\delta_1$ and $\delta_2^\ast= (\hat{r}/r_2)\delta_2$, $\delta_T=\sum_{n=1}^3 \delta_i$, $\delta_i$ is the radial thickness of zone $i$, $r_1 = (r_w+r_\mathrm{skin})/2$, $r_2 = (r_\mathrm{skin} + r_\mathrm{obs})/2$, and $\hat{r} = (r_w + r_\mathrm{obs})/2$. This approximate approach follows the work of \citet{shapiro1998} for using the equivalent hydraulic conductivity approach to account for simple heterogeneity. It is based on the simplifying assumption of a piecewise linear head distribution in the skin and formation. It follows directly from an application of mass conservation and Darcy's law in a radial (cylindrical) flow system. The result may also be obtained using a centered finite difference approximation of the hydraulic gradient at $r_1$ and $r_2$ for a head distribution given by Theim equation.

\subsection{Observation well storage \& inertial effects}
The column of water in the observation well oscillates in response to a source well perturbation. It is reasonable to assume that the effective weight of the water column in the observation well controls its head response and damping of the oscillations. Mass balance in the manner of \citet{blackkipp1977} and momentum balance \citep{kipp1985, butler2004} in the observation well account for wellbore storage and inertial effects. In non-dimensional form, the momentum balance equation is given by
\begin{equation}
\frac{\mathrm{d}^2 s_{D,\mathrm{obs}}}{\mathrm{d} t_D^2} + \gamma_{D,\mathrm{o}} \frac{\mathrm{d} s_{D,\mathrm{obs}}}{\mathrm{d} t_D} + \omega_{D,\mathrm{o}}^2 \; s_{D,\mathrm{obs}} = \omega_{D,\mathrm{o}}^2 \langle s_D \rangle
\label{eqn:dim-obs-well-storage}
\end{equation}
where $\gamma_{D,\mathrm{o}}$ is dimensionless observation well damping coefficient and $\omega_{D,\mathrm{o}}$ is dimensionless observation well characteristic frequency. where $s_{D,\mathrm{obs}}$ is the dimensionless observation well response, $\langle s_D \rangle$ is the depth-averaged dimensionless formation response across the observation interval. It follows from \citet{butler2004} that $\gamma_{D,\mathrm{o}} = T_c 8\nu L_\mathrm{obs}/(L_{e,\mathrm{obs}}r_{c,\mathrm{obs}}^2)$, $\omega_{D,\mathrm{o}} = T_c\sqrt{g/L_{e,\mathrm{obs}}}$, where $L_{e,\mathrm{obs}}$ and $L_\mathrm{obs}$ are the characteristic length scales for observation well inertial effects. Here we estimate $\gamma_\mathrm{o}$ and $\omega_\mathrm{o}$ through $L_\mathrm{obs}$ and $L_{e,\mathrm{obs}}$ from observation well data. Applying the Laplace transform and solving for $s_\mathrm{obs}$ gives
\begin{equation}
\overline{s}_{D,\mathrm{obs}} = \overline{\psi_2}(p) \left\langle \overline{s}_D \left(r_D,p \right) \right\rangle,
\end{equation}
where $\overline{\psi}_2(p) = \omega_{D,\mathrm{o}}^2/(p^2 + p \gamma_{D,\mathrm{o}} + \omega_{D,\mathrm{o}}^2)$, 
\begin{equation}
\label{eqn:s_obs}
\langle \hat{\overline{s}}_D \rangle = \frac{1}{b_{D,\mathrm{o}}} \int_{d_{D,\mathrm{o}}}^{l_{D,\mathrm{o}}} \hat{\overline{s}}_D (a_i,p,z_D) \;\mathrm{d}z_D,
\end{equation}
and $l_{D,\mathrm{o}}=l_\mathrm{o}/B$ and $d_{D,\mathrm{o}}=d_\mathrm{o}/B$ are the dimensionless depths to the top and bottom of the observation well interval from the water-table. Upon inverting the Laplace transform, one obtains
\begin{equation}\label{eqn:obsresponse}
s_{D,\mathrm{obs}} = \int_0^{t_D} \psi_2(t_D-\tau) \left\langle s_D \left(r_D,\tau \right) \right\rangle \;\mathrm{d}\tau
\end{equation}
with $\psi_2(t) = \mathcal{L}^{-1} \left\{\overline{\psi}_2(p) \right\}$. Equation \ref{eqn:obsresponse} is the solution accounting for observation well inertial effects. It is used in the subsequent analysis to estimate hydraulic parameters.

\section{Model Application to Cross-hole Slug Test Data}
\begin{figure}[h] % Figure 2
  \centering
  \includegraphics[width=0.5\textwidth]{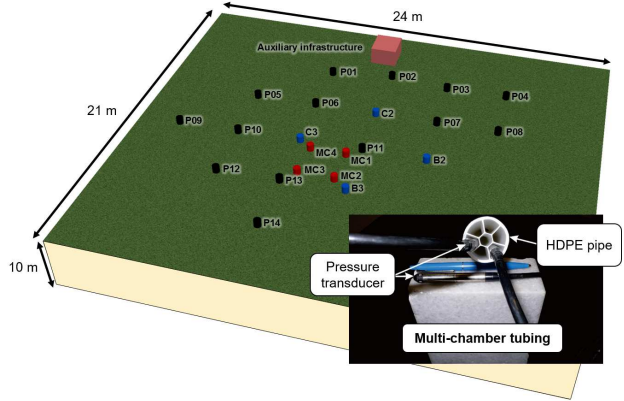}\\(a)\\ \includegraphics[width=0.5\textwidth]{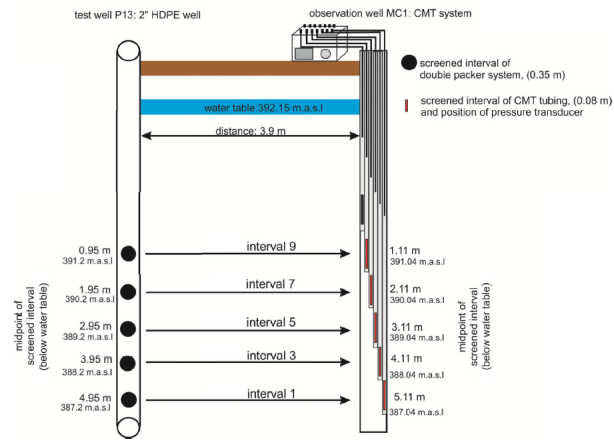}\\(b)\\
\caption{\label{fig:setup} (a) Multi-well layout and (b) example experimental system setup for cross-hole slug tests at the Widen site, Switzerland. For the example shown, data from interval $i$ is denoted P13-MC1-$i$}
\end{figure}

The model described above is applied to observations collected in a series of multi-level cross-hole pneumatic slug tests performed in June 2013 at the Widen site in north-east Switzerland. The site is on the floodplain of the Thur River, a tributary of the Rhine river \citep{diem2010}. The multi-well layout of the test site is depicted schematically in Figure \ref{fig:setup}(a). The wells used in the experiments are completed in an unconfined sand and gravel aquifer with a saturated thickness of 5.8~m. The aquifer is quaternary post-glacial sediment underlain by an aquitard of low permeability lacustrine sediment comprising fine silt and clay \citep{diem2010, coscia2011}. It is overlain with alluvial loam that constitutes the top soil. The aquifer itself can be further subdivided into a silty sand top layer underlain with silty gravel and a sand layer to a thickness of about 7 m \citep{diem2010}. The source well is screened across the entire saturated thickness (see Figure \ref{fig:setup}(a)). Straddle packers were used to sequentially isolate discrete intervals in the source well. The pressure responses were recorded in three observation wells, which were equipped with a Continuous Multichannel Tubing (CMT) system \citep{einarson2002} in which pressure transducers were installed. This system was originally designed for multi-level sampling. It consists of a PVC pipe with seven continuous separate channels or chambers (inner diameter 0.014~m), which are arranged in a honeycomb structure. Each individual chamber has a 0.08~m long slot covered with a sand filter and allows for hydraulic contact with the formation.

\subsection{Experimental procedure}
The cross-hole pneumatic slug tests were initiated by applying gas pressure to the water column in a chosen interval, then releasing the gas pressure through an outflow valve to provide the instantaneous initial slug perturbation. A double-packer system straddling the test interval ($b_s = 0.35$~m) was used with the pneumatic slug applied through a smaller tubing ($r_c = 1.55\times 10^{-2}$~m). The source well used in these tests was well P13, with a wellbore radius of $r_w = 3.15\times 10^{-2}$~m. The dissipation of the slug was monitored with a pressure transducer in the source well positioned at the top of the water column above the test interval.

The data considered here was obtained in three observations wells labelled MC1, MC2, and MC4 (in Figure \ref{fig:setup}(a)) and located at radial distances of 3.9, 2.9, and 2.8~m, respectively, from the source well. The responses at multiple vertical positions in each observation well were monitored with pressure transducers in a seven-channel CMT system with screen intervals of $b_\mathrm{o} = 8\times 10^{-2}$~m. Each channel in the CMT system has an equivalent radius of $r_{c,\mathrm{o}}=6.5\times 10^{-3}$~m; installation of a pressure transducer in these channels reduces their effective radii (and effective wellbore storage) significantly. The CMT system allows for simultaneous monitoring of the response at seven vertical positions for each slug test. Pressure responses were recorded at a frequency of 50~Hz (every 0.02~s) for a period of about 20 seconds from slug initiation using miniature submersible level transmitters MTM/N 10 manufactured by STS Sensor Technik in Switzerland. The housing diameter of 0.39 inches allowed for pressure measurements in small diameter (1/2 inch) monitoring wells, stand pipes and bore holes. The stainless steel construction and integral polyurethane cable is ideal for long term installation. The transducer cable is reinforced with Kevlar to avoid elongation in deep boreholes. The experiments reported herein were performed in shallow wells and over a relatively short duration to make cable elongation is negligible.

Only data from the observation intervals at approximately the same vertical position as the source-well test interval are analysed here because of their favourable signal-to-noise ratio (SNR). Data from ports not directly in line with the tested interval showed significant decay for the magnitudes of the perturbation used in the field tests. Transducers with greater precision and accuracy or larger source well perturbation are needed to obtain analysable responses in such ports. A schematic of the experimental setup for tests between wells P13 and MC1 is shown in Figure \ref{fig:setup}(b). 

\subsection{Observation well data}
\begin{figure}[h] % Figure 3
\includegraphics[width=0.48\textwidth]{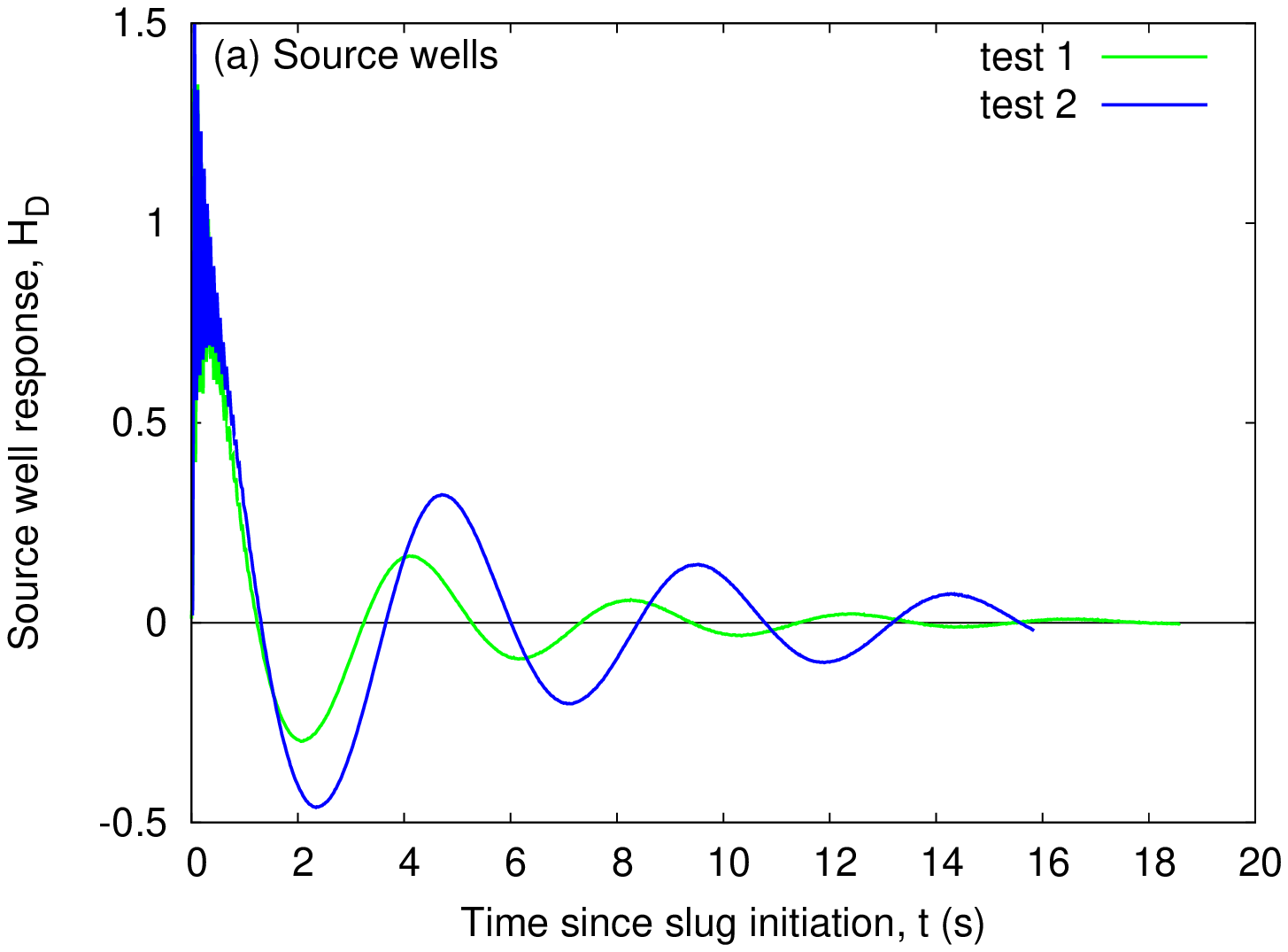} \includegraphics[width=0.48\textwidth]{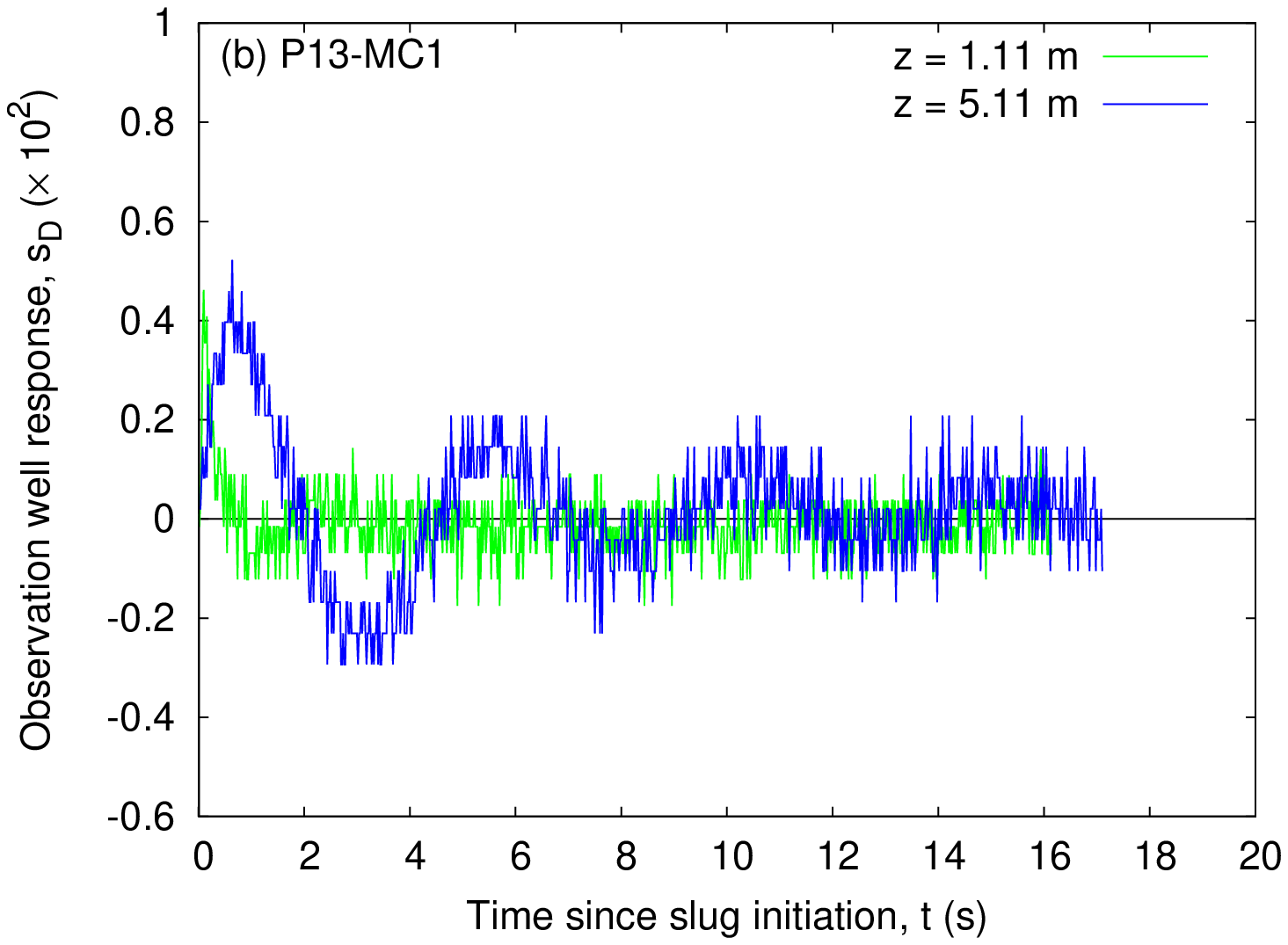}\\
\includegraphics[width=0.48\textwidth]{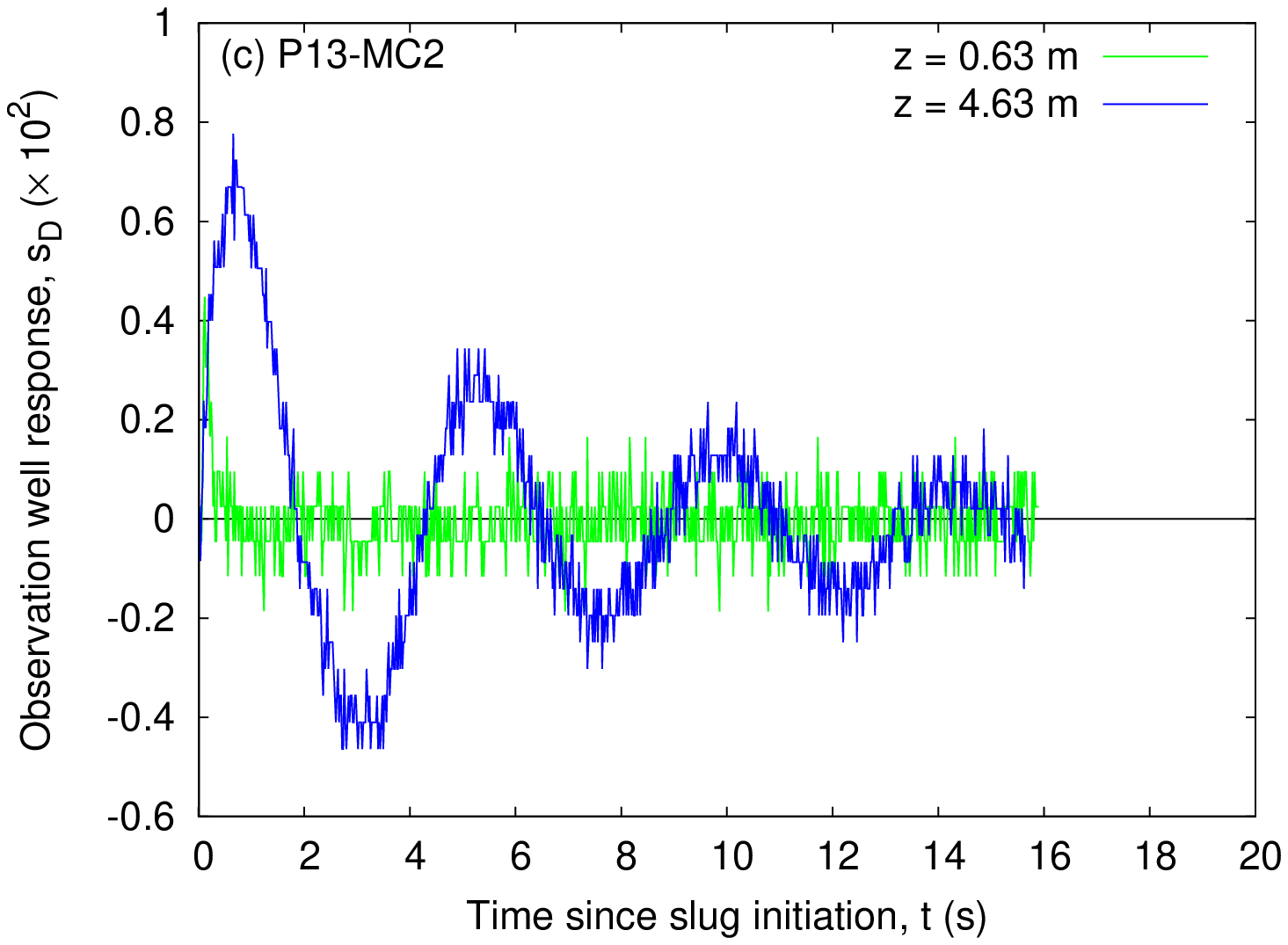} \includegraphics[width=0.48\textwidth]{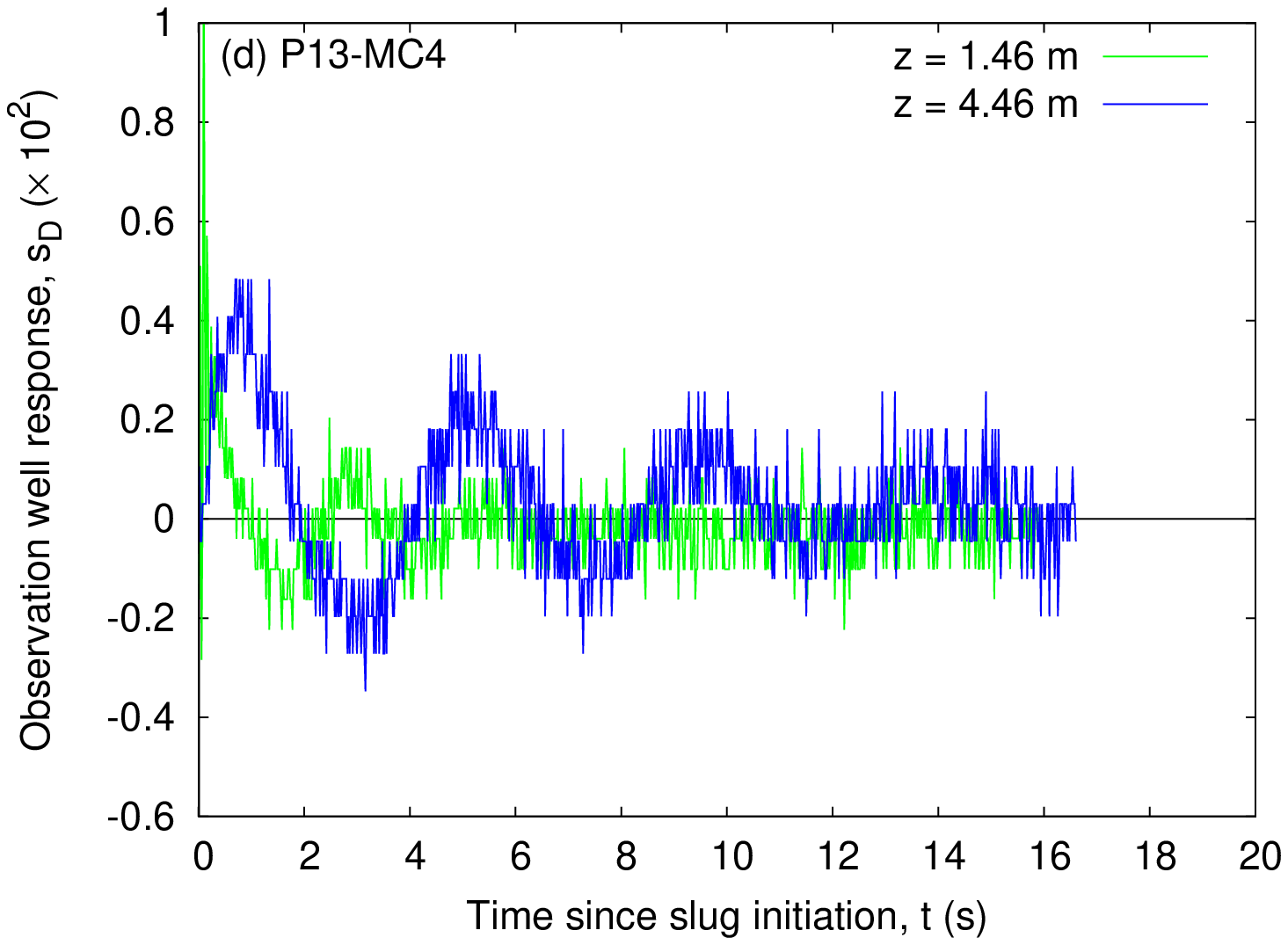}
\caption{\label{fig:typical-obs} Typical (a) source and (b-d) observation well responses measured during cross-hole slug tests. Observation well data show increasing damping when approaching the watertable for all three profiles.}
\end{figure}

The typical slug test responses observed during tests at the Widen site are shown in Figure \ref{fig:typical-obs}. The plots in Figure \ref{fig:typical-obs}(a) are the source well responses, and those in (b) are the corresponding responses in an observation well about 3~m radially from the source well. The results clearly show damped oscillations generated in the source well are measurable in an observation well a few meters away. Comparing the results in (a) and (b) also shows the maximum amplitude of the signal decays about two orders of magnitude from the source to the observation well, which decreases the SNR.

The observation well response pairs generally are increasingly damped moving towards the water-table, even when the initial displacements from the equilibrium position are comparable. This is evident in the data from all three profiles shown in Figure \ref{fig:typical-obs}, where observation well data collected closer to the water-table appear to be more damped than those at greater depths. Measurable observation well displacements are still obtainable near the water-table (i.e., interval 9 in Figure~\ref{fig:setup}(b)). The configuration of the equipment made it physically impossible to record the response at the water-table. Placing a pressure transducer at the water-table would be useful to confirm the appropriate type of boundary condition to represent the water-table. While \cite{malama2011} and the modified model presented here use the linearised kinematic water-table  representation,  \cite{hyder1994} use a constant-head boundary condition to represent the water-table.

\subsection{Parameter estimation}
The modified model was used to estimate model parameters from data collected in observation wells during the tests at the Widen site. For the present study, to reduce the number of estimated parameters, it is sufficient to assume the aquifer is isotropic ($K_r = K_z = K$), and the skin conductivities of the source and observation wells are equal ($K_1 = K_3 = K_\mathrm{skin}$). Using the non-linear optimization software PEST \citep{doherty2010, doherty15}, we estimated skin hydraulic conductivity ($K_\mathrm{skin}$), formation hydraulic conductivity ($K$), specific storage ($S_s$), and the length parameters $L$ and $L_e$ that characterize the source and observation well damping coefficients and frequencies. It is typical to compute $L$ and $L_e$ using the formulas \citep{butler2002, kipp1985, zurbuchen2002}
\begin{equation}
\label{eqn:L}
L = d + \frac{b}{2} \left(\frac{r_c}{r_w}\right)^4,
\end{equation}
and
\begin{equation}
\label{eqn:Le}
L_e = L + \frac{b}{2} \left(\frac{r_c}{r_w}\right)^2.
\end{equation}
The values of $L$ and $L_e$ computed with these formulas were used as initial guesses during the parameter estimation procedure. The parameters $L_{e,\mathrm{obs}}$ and $L_\mathrm{obs}$, which determine the frequency and damping coefficient of the observation well were also estimated with initial guesses determined similarly. The non-linear optimization software PEST \citep{doherty2010, doherty15} was used to estimate the optimal parameters and the model parameter sensitivity at the optimal solution.

The fit of the model to observed cross-hole responses was very sensitive to the time of the initial observation (i.e., the syncing of the clocks at the source and observation wells). Initially it was difficult to get model/data agreement to both early and late-time data without assigning non-physical parameter values. Estimating a modest time shift (off-set) for each test greatly improved model fits to the data. Estimated observation data time delays were between 4 and 6 tenths of a second, which is a permissible time off-set between two synced transducer clocks.

\begin{figure}[h] % Figure 4
  %\centering
  \begin{tabular}{ccc}
  \includegraphics[width=0.32\textwidth]{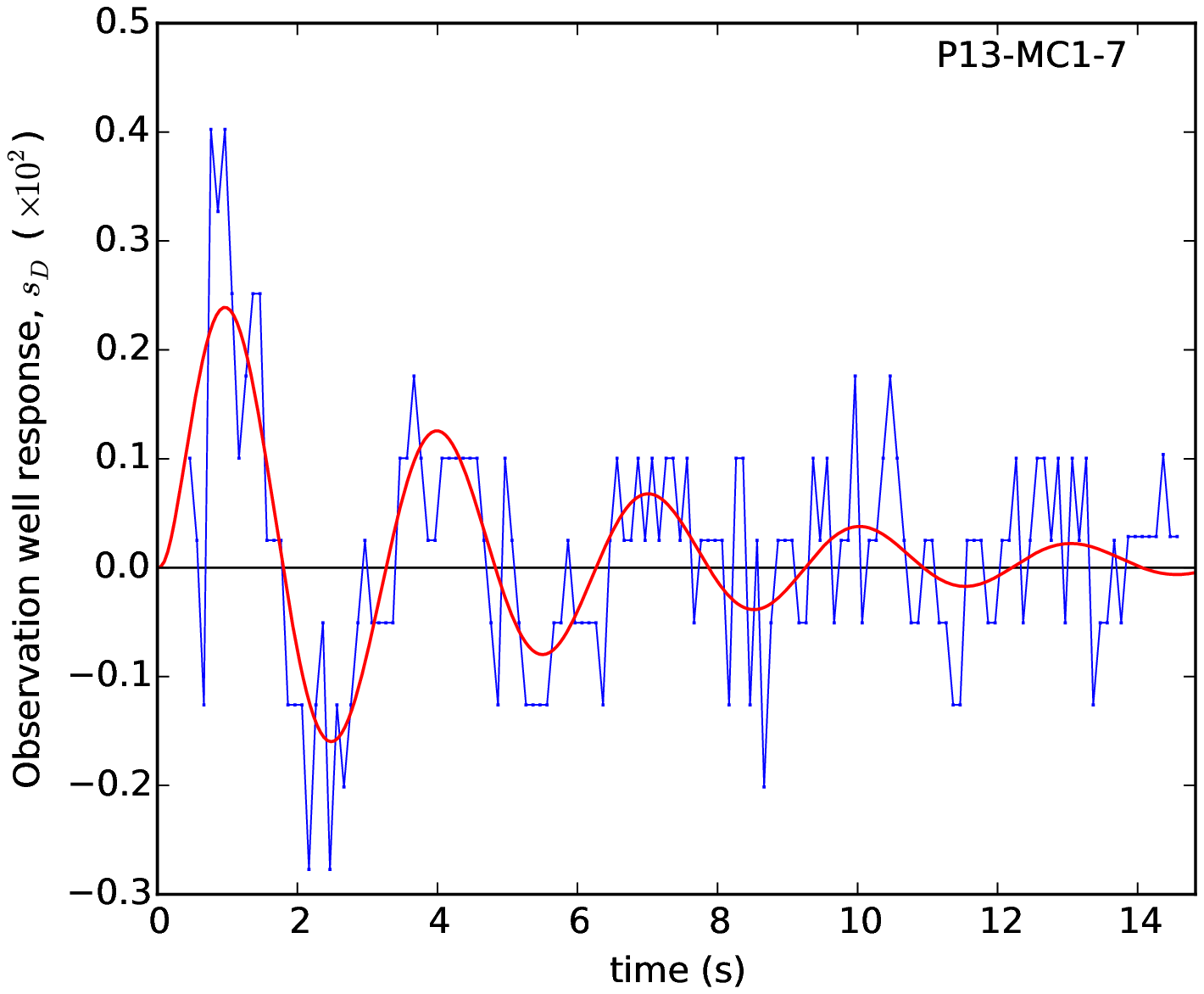} & \includegraphics[width=0.32\textwidth]{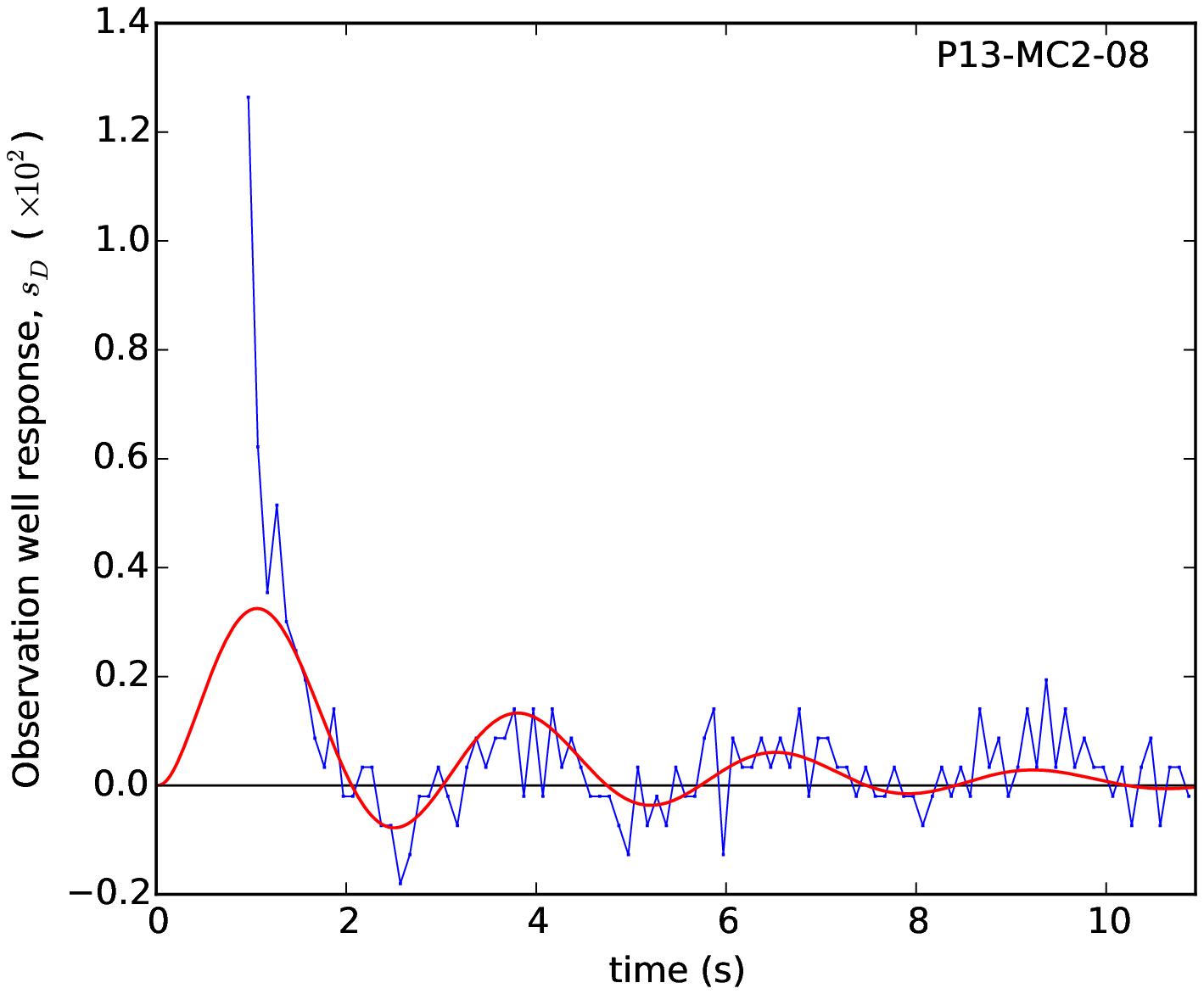} & \includegraphics[width=0.32\textwidth]{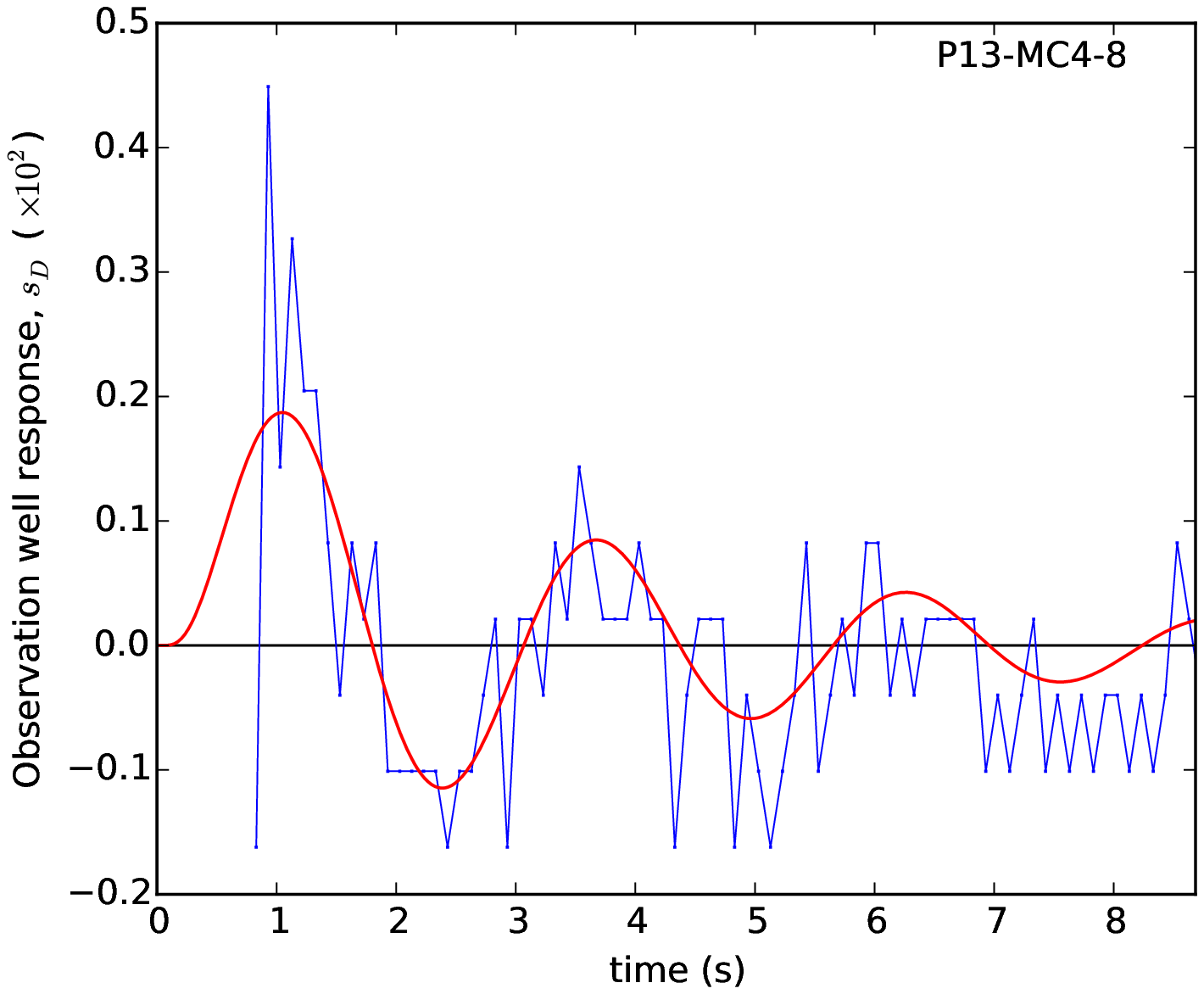}\\

  \includegraphics[width=0.32\textwidth]{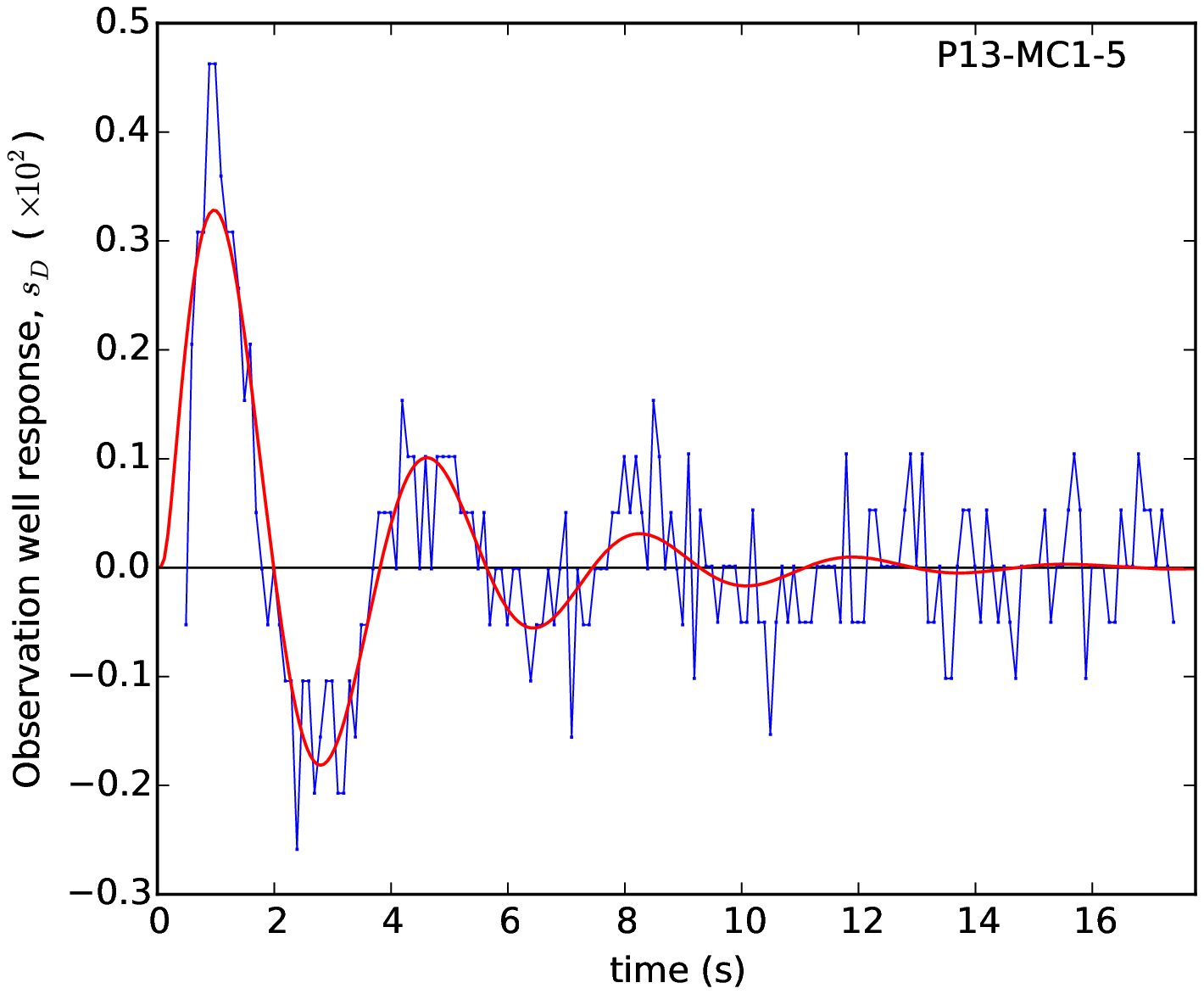} & \includegraphics[width=0.32\textwidth]{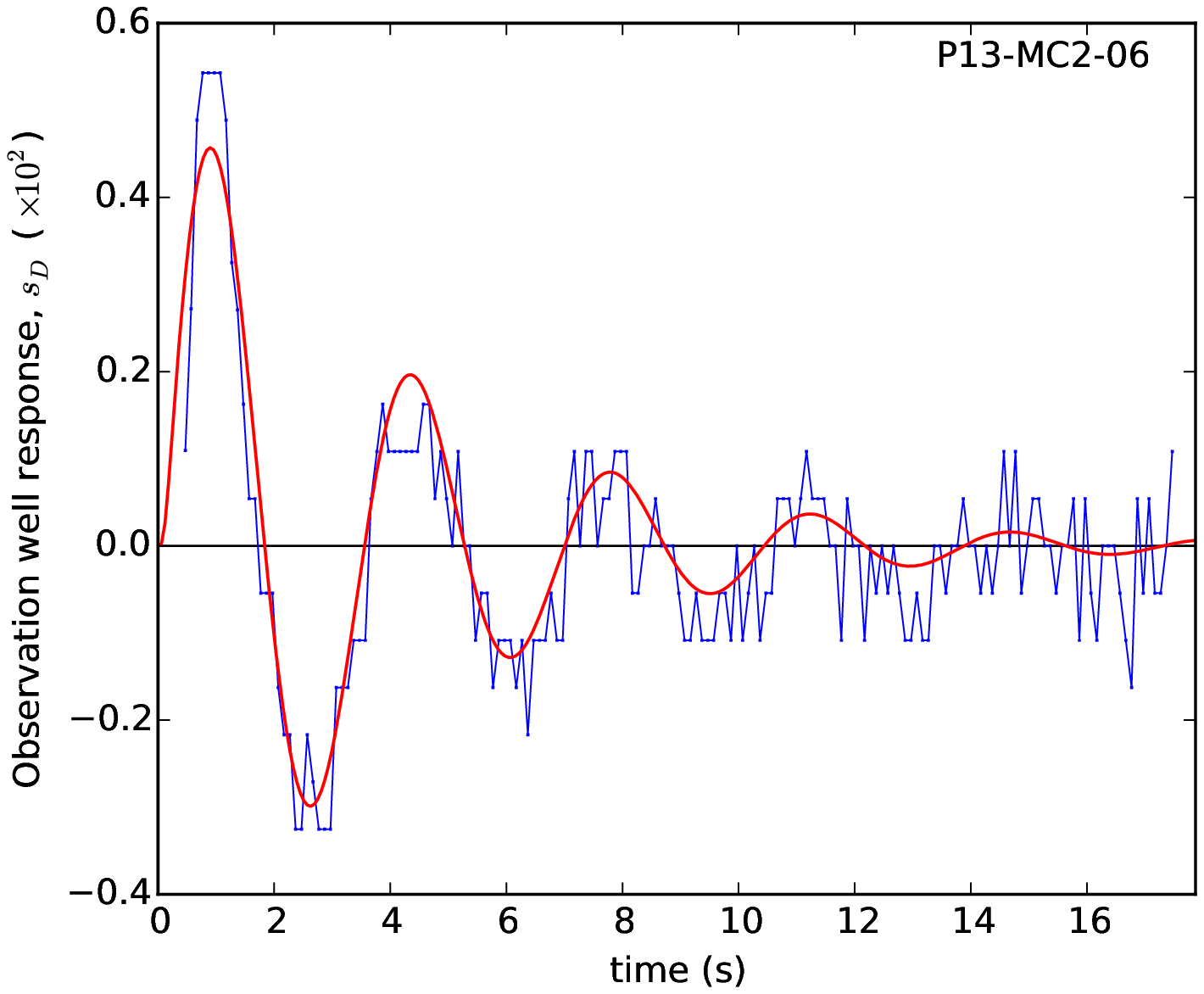} & \includegraphics[width=0.32\textwidth]{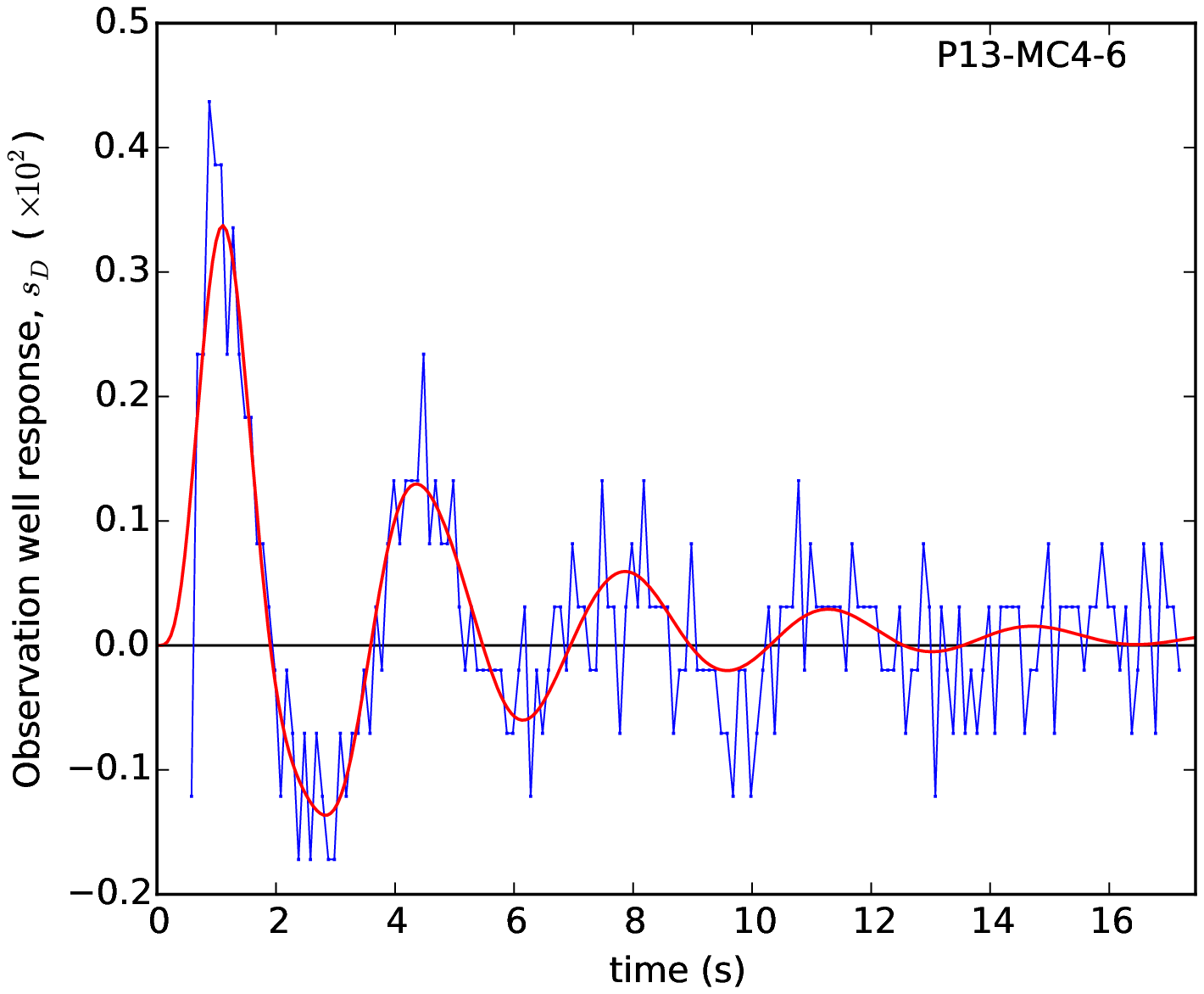}\\

  \includegraphics[width=0.32\textwidth]{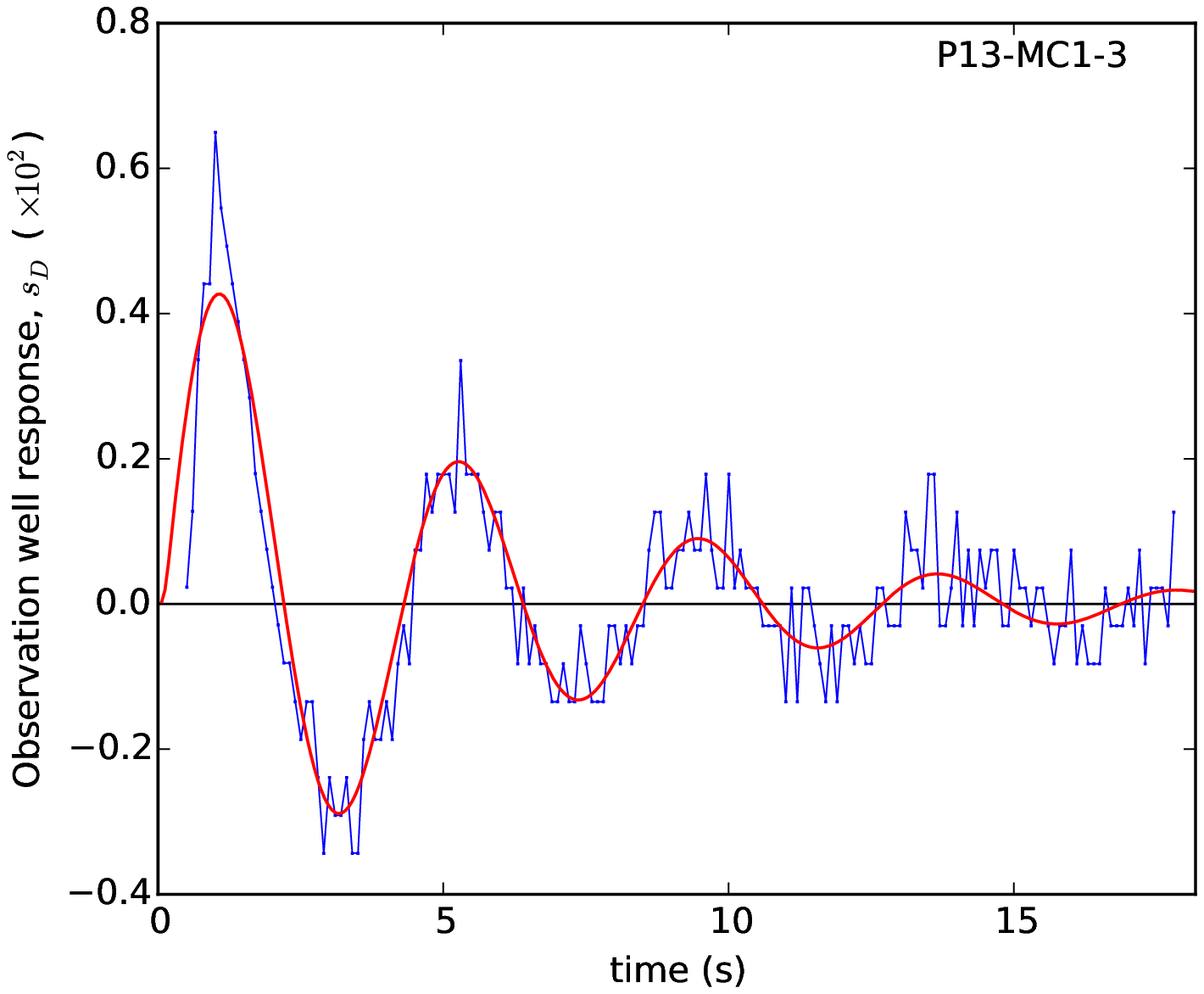} & \includegraphics[width=0.32\textwidth]{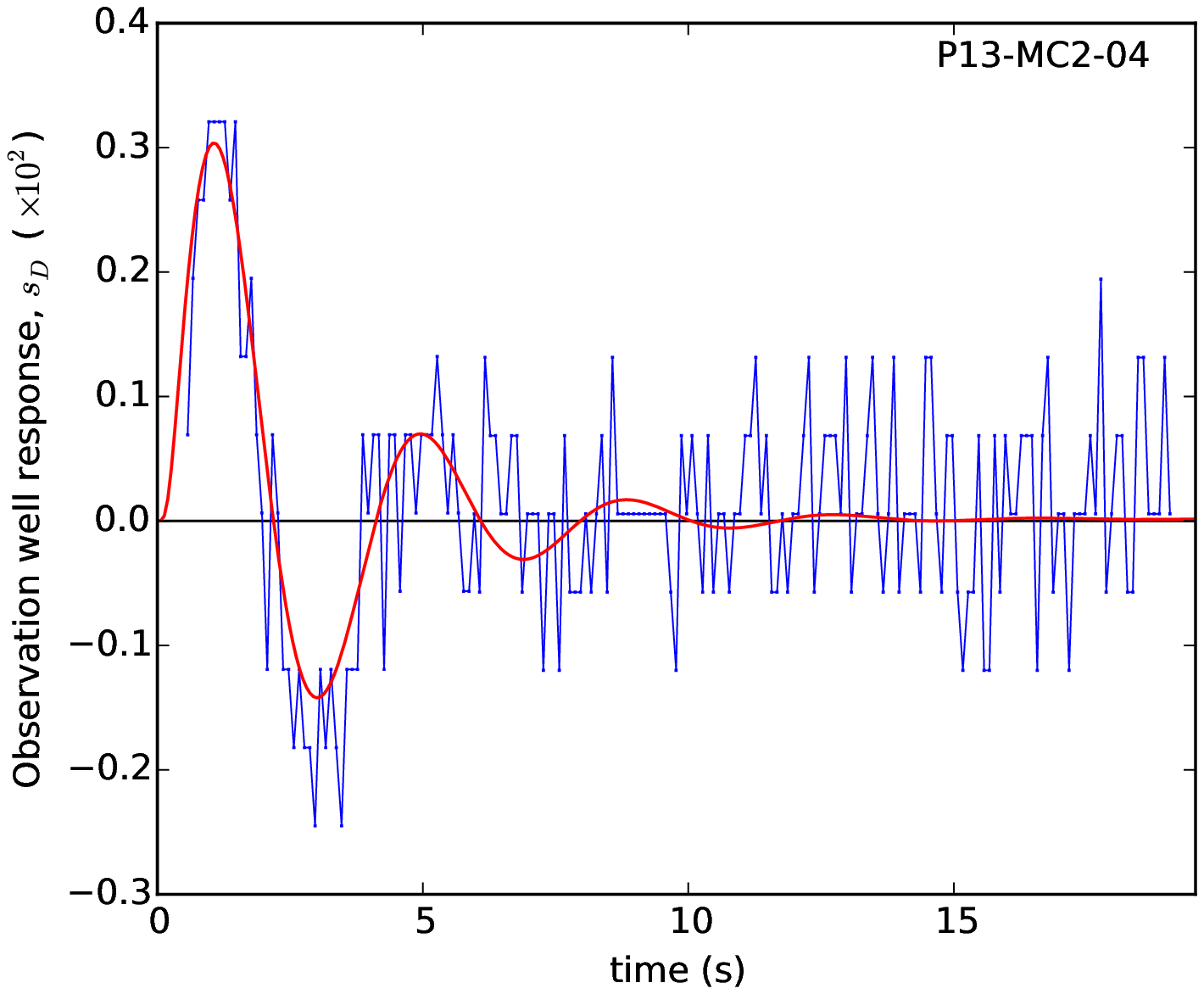} & \includegraphics[width=0.32\textwidth]{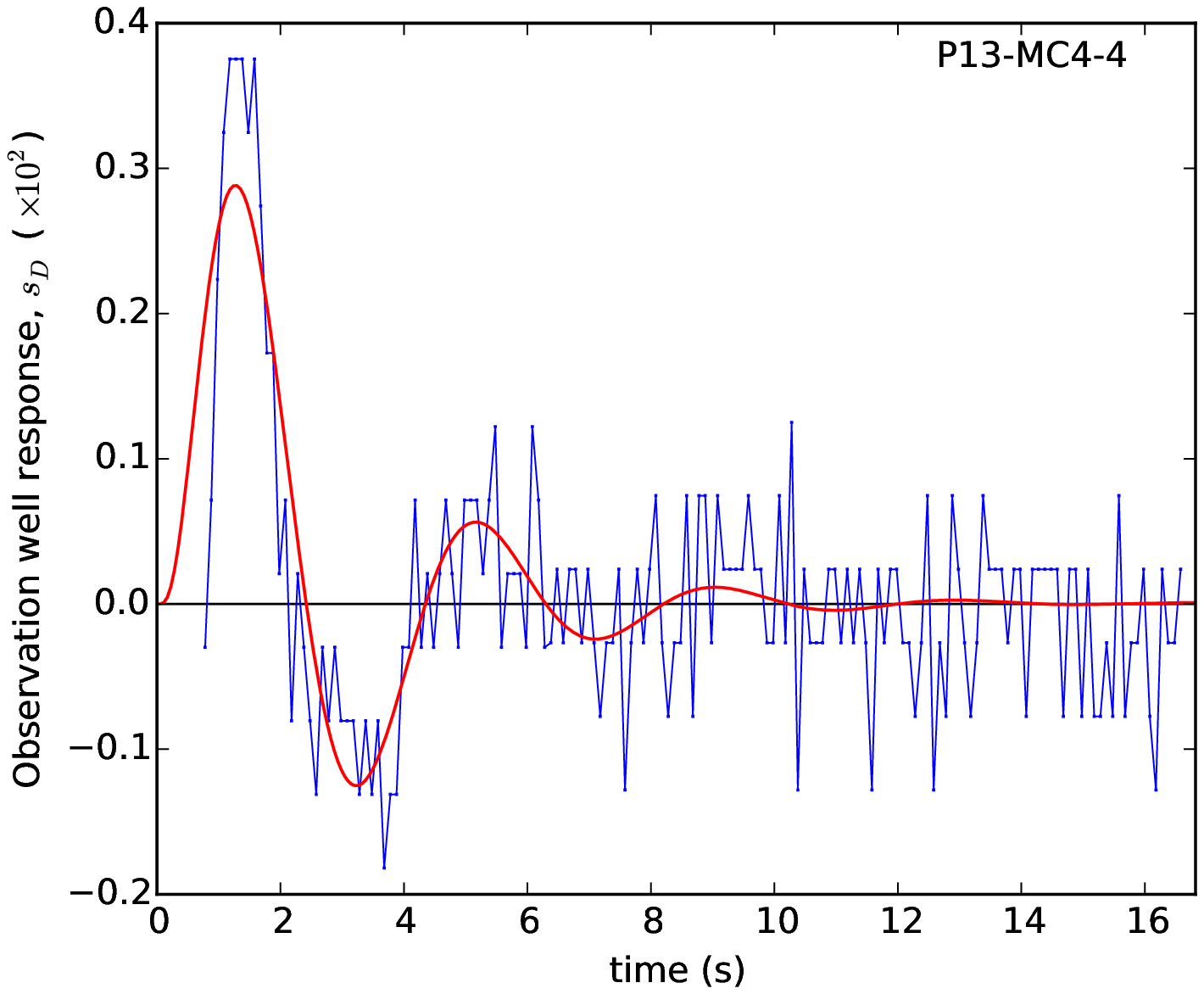}\\

  \includegraphics[width=0.32\textwidth]{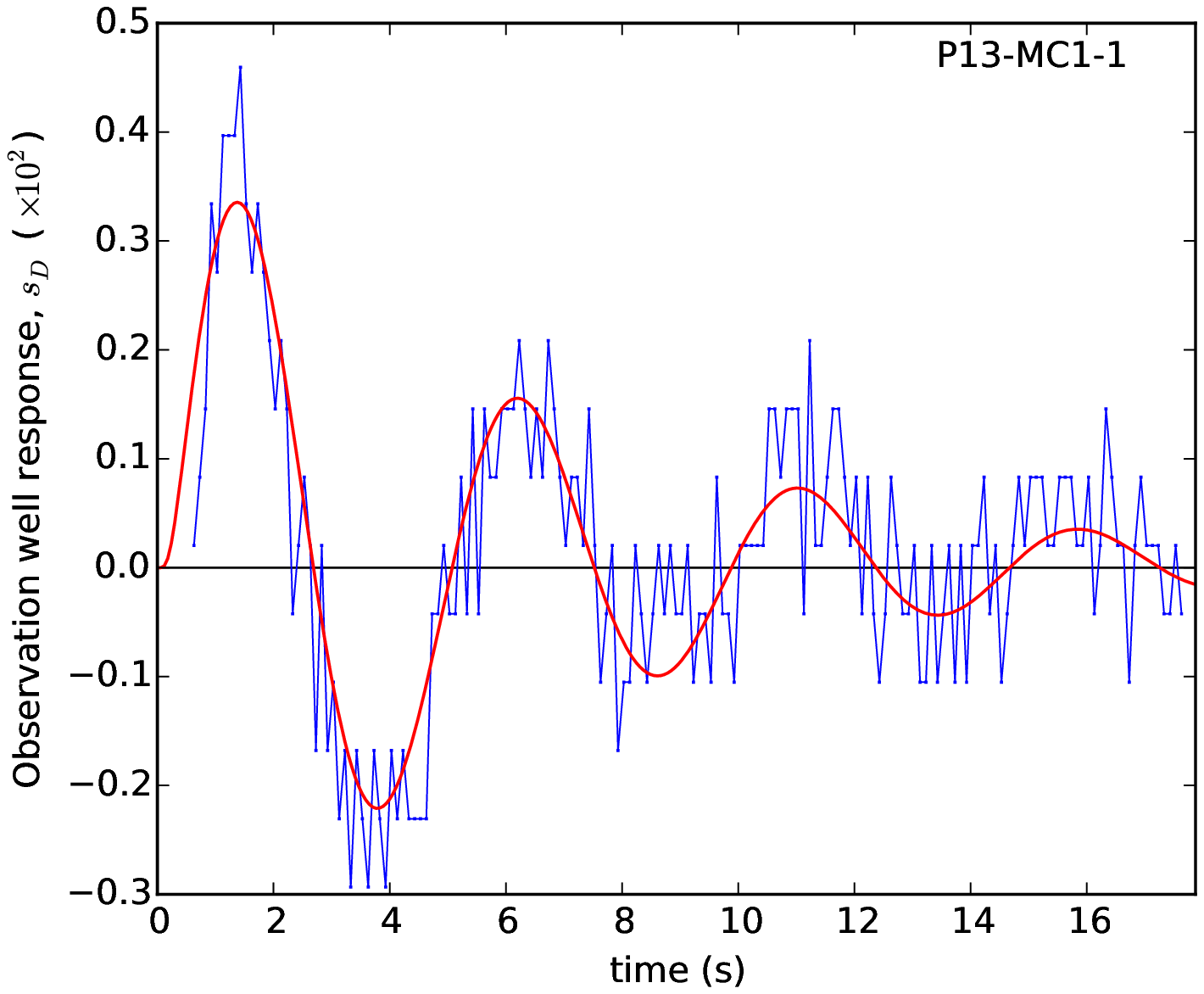} & \includegraphics[width=0.32\textwidth]{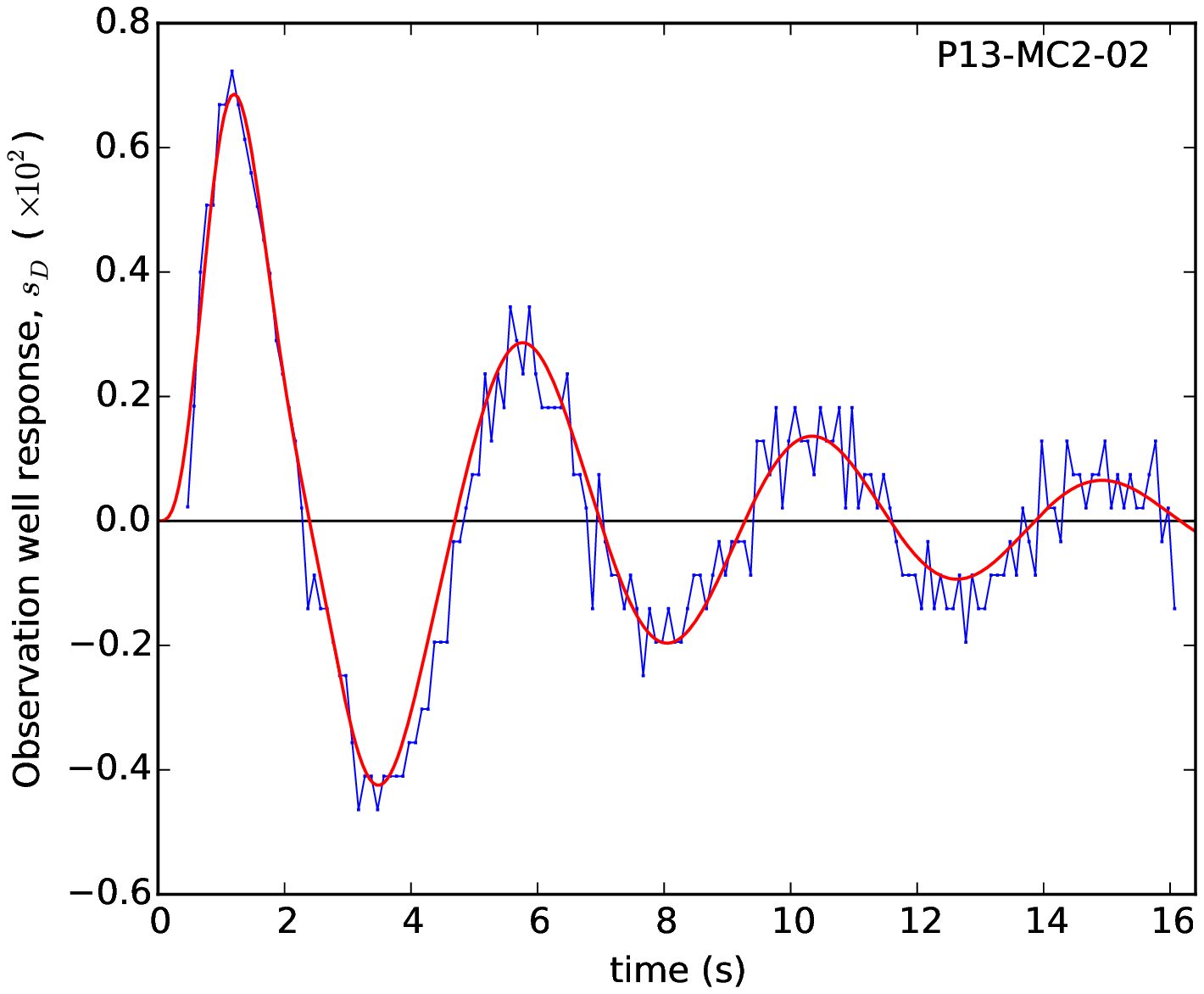} & \includegraphics[width=0.32\textwidth]{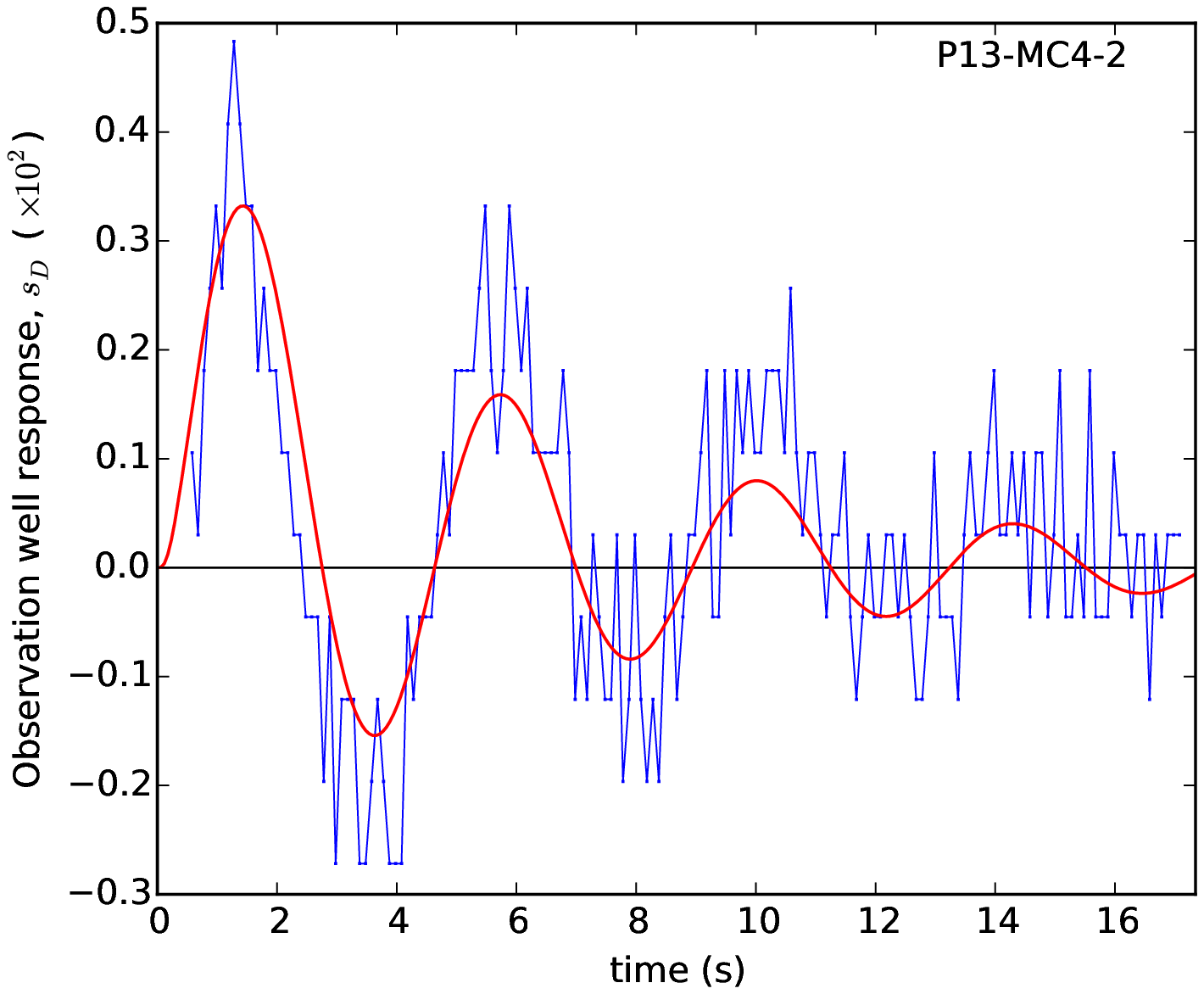}
  \end{tabular}
  \caption{\label{fig:modelfits} Model fits to cross-hole slug test data collected along vertical profiles in three observation wells at the Widen Site, Switzerland. The columns correspond to profiles in observation wells 1, 2, and 4.}
\end{figure}

PEST-estimated parameters are summarized in Table \ref{tab:params}. A subset of the complete dataset (25\% of the 50 Hz data stream) was used in the PEST optimization; this subset is shown in Figure~\ref{fig:modelfits}. The corresponding model fits to observation well data are shown in Figure \ref{fig:modelfits}. The relatively large average value of skin conductivity (averaging $K_\mathrm{skin} = 8.5\times 10^{-2}$~m/s) estimated from tests is consistent with a disturbed zone resulting from well installation by direct-push. The technology uses a hydraulic hammer supplemented with weight of the direct-push unit to push down drive rods to the desired depth of the projected well. The well casing  is then lowered into the drive rods (inner diameter: 0.067~m, outer diameter 0.083~m). By retracting the drive rods, the formation is allowed to collapse back against the casing. The negative skin estimates ($K_\mathrm{skin}$ greater than formation $K$) are indicative of formation collapse due to material bridging resulting in a disturbed zone around the well casing. Skin conductivity estimation variances range from $10^{-2}\;\mathrm{m^2/s^2}$ for low noise data to $10^3\;\mathrm{m^2/s^2}$ noisy data and are indicative of dependence of estimation uncertainties on measurement errors.

\begin{table}[ht] % Table 2
\caption{\label{tab:params}PEST-estimated model parameters.}
\centering
\begin{tabular}{lcccccccccc}
\hline 
     & $K$ & $K_\mathrm{skin}$ & $S_s$ & $S_y$ & $L$ & $L_e$ & $L_\mathrm{obs}$ & $L_{e,\mathrm{obs}}$\\
Test & [$\mathrm{m \cdot s^{-1}}$] & [$\mathrm{m \cdot s^{-1}}$] & [$\mathrm{m^{-1}}$] & [-] & [m] & [m] & [m] & [m] \\
\hline
P13-MC1-1 & $7.81\times 10^{-4}$ & $2.27\times 10^{-1}$ & $3.39\times 10^{-5}$ & 0.037 & 1.90 & 5.71 & 4.07 & $1.87\times 10^{-2}$\\
P13-MC1-3 & $8.85\times 10^{-4}$ & $1.07\times 10^{-1}$ & $1.25\times 10^{-5}$ & 0.40 & 1.18 & 4.31 & 2.39 & $1.80\times 10^{-2}$\\
P13-MC1-5 & $7.70\times 10^{-4}$ & $2.21\times 10^{-1}$ & $1.70\times 10^{-5}$ & 0.36 & 2.53 & 3.21 & 3.10 & $1.76\times 10^{-2}$\\
P13-MC1-7 & $1.28\times 10^{-3}$ & $1.02\times 10^{-2}$ & $3.85\times 10^{-5}$ & 0.018 & 0.23 & 2.26 & 11.4 & $9.74\times 10^{-2}$\\
P13-MC1-9 & $1.48\times 10^{-3}$ & $6.42\times 10^{2}$ & $2.79\times 10^{-8}$ & 0.001 & 2.95 & 0.83 & 8.83 & $6.04\times 10^{-2}$\\
\hline
P13-MC2-2 & $7.67\times 10^{-4}$ & $1.73\times 10^{-1}$ & $2.76\times 10^{-5}$ & 0.40 & 1.07 & 5.07 & 3.92 & $5.73\times10^{-1}$ \\
P13-MC2-4 & $1.36\times 10^{-3}$ & $2.15\times 10^{-1}$ & $5.11\times 10^{-5}$ & 0.04 & 8.01 & 3.66 & 4.63 & $4.84\times10^{-2}$\\
P13-MC2-6 & $1.08\times 10^{-3}$ & $6.42\times 10^{-2}$ & $1.95\times 10^{-5}$ & 0.40 & 1.38 & 2.91 & 2.55 & $1.79\times 10^{-2}$\\
P13-MC2-8 & $2.22\times 10^{-3}$ & $4.44\times 10^{-1}$ & $3.06\times 10^{-5}$ & 0.001 & 2.26 & 1.80 & 5.34 & $1.64\times 10^{-3}$\\
\hline
P13-MC4-2 & $1.60\times 10^{-3}$ & $3.17\times 10^{-1}$ & $7.41\times 10^{-5}$ & 0.005 & 2.39 & 4.81 & 6.42 & $1.35\times 10^{-2}$\\
P13-MC4-4 & $5.27\times 10^{-4}$ & $3.37\times 10^{-1}$ & $9.16\times 10^{-5}$ & 0.40 & 4.56 & 3.49 & 9.69 & $1.38\times 10^{-2}$\\
P13-MC4-6 & $3.79\times 10^{-3}$ & $2.04\times 10^{-2}$ & $7.22\times 10^{-5}$ & 0.001 & 3.24 & 2.82 & 3.09 & $4.85\times 10^{-1}$\\
P13-MC4-8 & $1.46\times 10^{-3}$ & $6.15\times 10^{-2}$ & $1.80\times 10^{-4}$ & 0.40 & 0.78 & 1.76 & 9.24 & $3.48\times 10^{-2}$\\
\hline
\end{tabular}
\end{table}
\FloatBarrier

\FloatBarrier

Drilling logs and previous hydrogeophysical investigations at the site \citep{lochbuhler2013, coscia2011} indicate a sand and gravel aquifer. The formation hydraulic conductivities estimated here are on the order of $10^{-4}$ to $10^{-3}$~m/s, and in general agreement with the findings from earlier studies at the site. \citet{coscia2011} report estimates of the order of $10^{-3}$ to $10^{-2}$~m/s from multiple pumping and single-well slug tests conducted at the site by \citet{diem2010}. The average values estimated here range from a low of $7.1 \times 10^{-4}$~m/s to a high value of $3.8 \times 10^{-3}$~m/s. These and estimates from earlier studies at the site are reasonable for unconsolidated well-sorted sand and gravel aquifers \citep{bear1972, fetter2001}. The vertical variability in the estimates is reflective of site heterogeneity. The objective of multi-level slug tests is to characterize such heterogeneity using a physically based flow model. It should be understood that the model used in this analysis was developed for a homogeneous but anisotropic aquifer. Its application to characterizing heterogeneity is thus limited and only approximate, with data collected at discrete depth intervals assumed to yield hydraulic parameter values associated with that interval. Estimation variances for formation hydraulic conductivity range in magnitude from $6\times 10^{-2}$ to $1.2\times 10^1\;\mathrm{m^2/s^2}$.

Estimates of specific storage, $S_s$, also show only modest variability and are generally of the order of $10^{-5}\;\mathrm{m^{-1}}$, with the largest value being about $10^{-5}\;\mathrm{m^{-1}}$ and the smallest $10^{-7}\;\mathrm{m^{-1}}$. The estimated values are indicative of poorly consolidated shallow alluvium, and variability may reflect uncertainty or non-uniqueness in the solution for this configuration and dataset. Estimates of $S_y$ were quite variable, with estimation variances of the order of $10^{-7}$ to $10^2$. Estimated values of 0.4 correspond to the upper bound during optimization. P13-MC1-1, 7, and 9 resulted in estimated $S_y$ values of a few percent, which are physically realistic for these types of sediments and for the linearized kinematic condition at the watertable. In this parameter estimation analysis no significant physical constraints where introduced on the objective function; the observations were allowed to freely constrain the estimates of model parameters. Estimates of the parameter $L_e$ from data are comparable to those predicted by equation (\ref{eqn:Le}). However, estimates of $L$ from data are consistently larger than the values predicted by (\ref{eqn:L}).

\subsection{Sensitivity analysis}
\begin{figure}[h] % Figure 5
\includegraphics[width=0.48\textwidth]{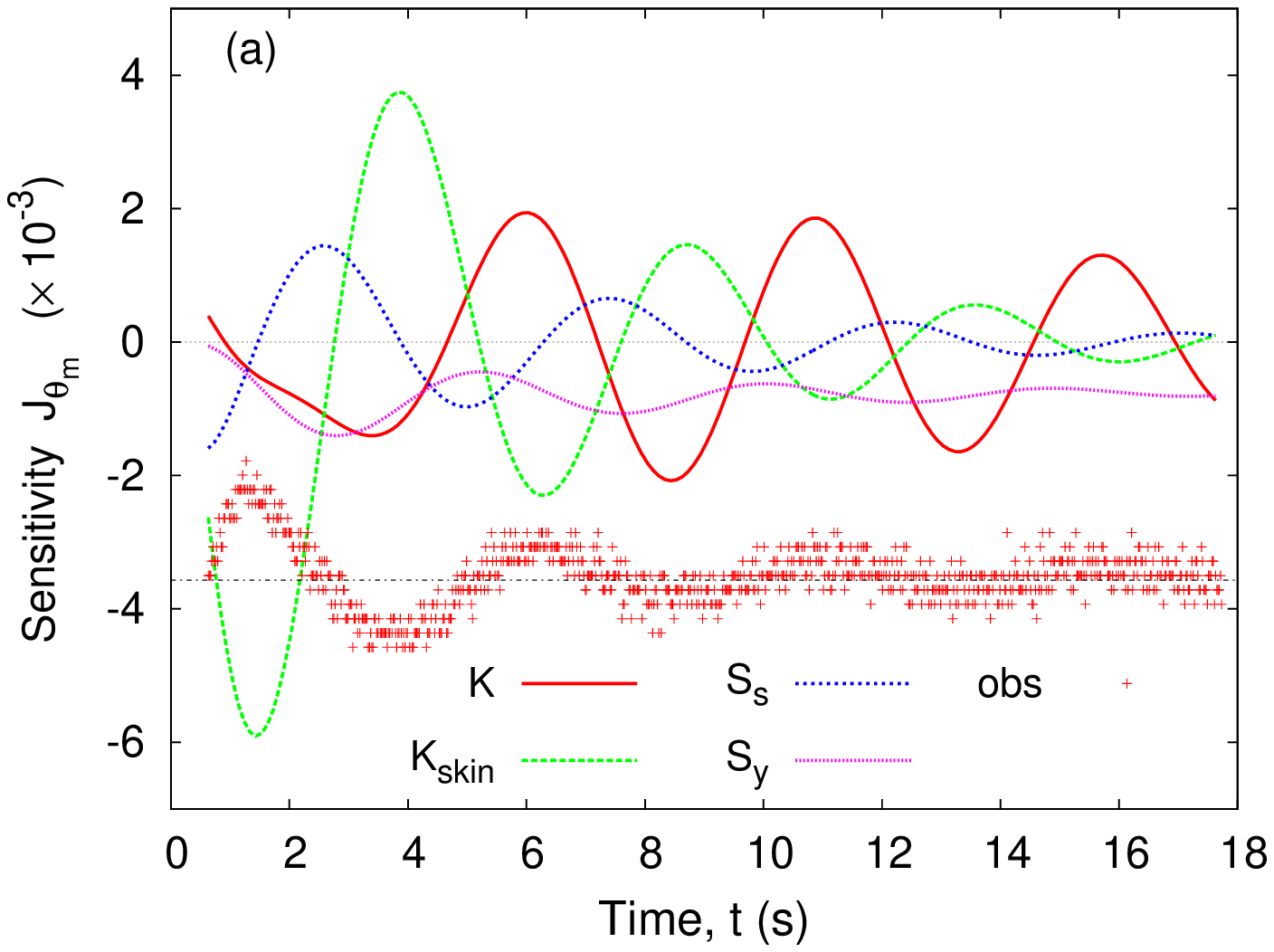}
\includegraphics[width=0.48\textwidth]{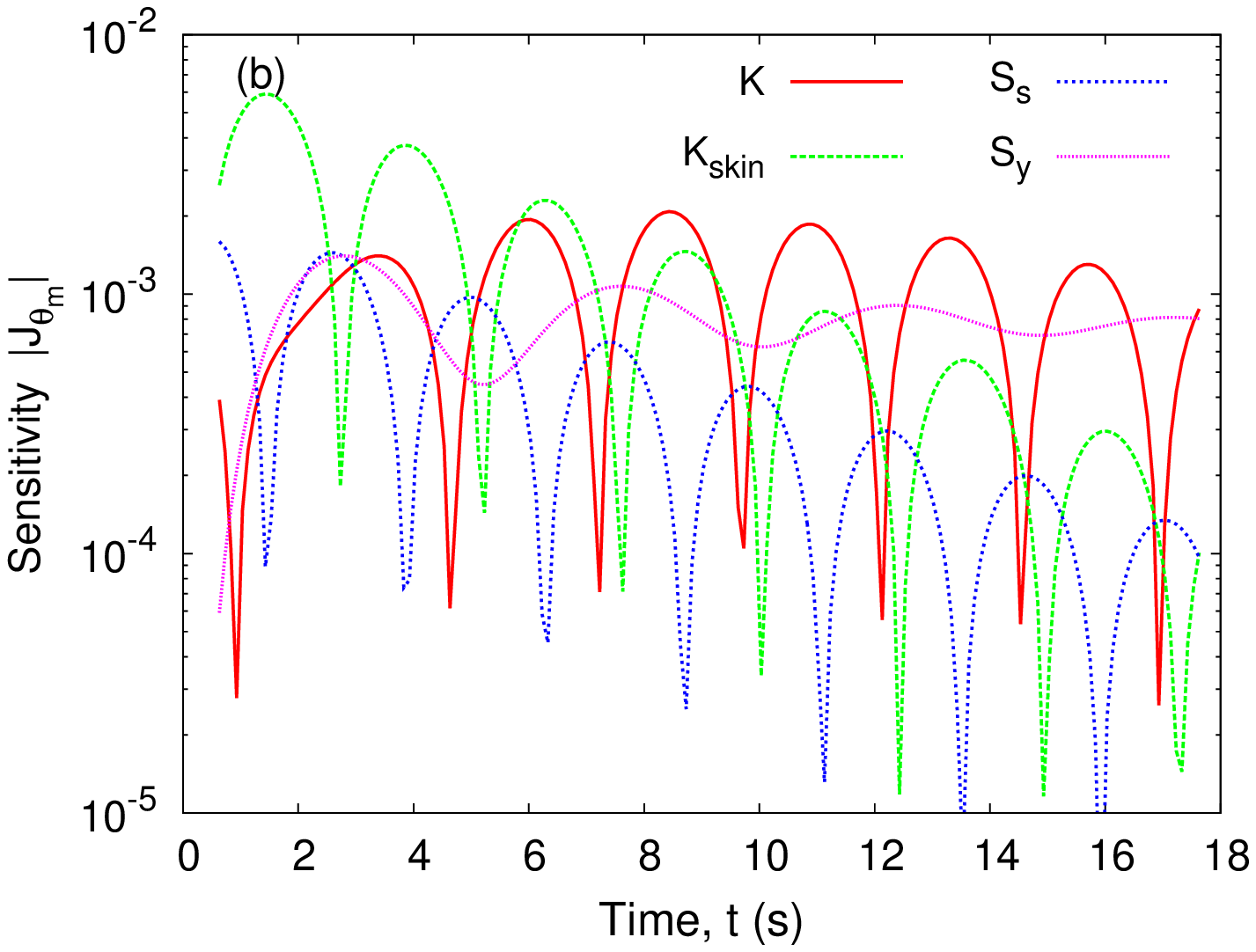}
\includegraphics[width=0.48\textwidth]{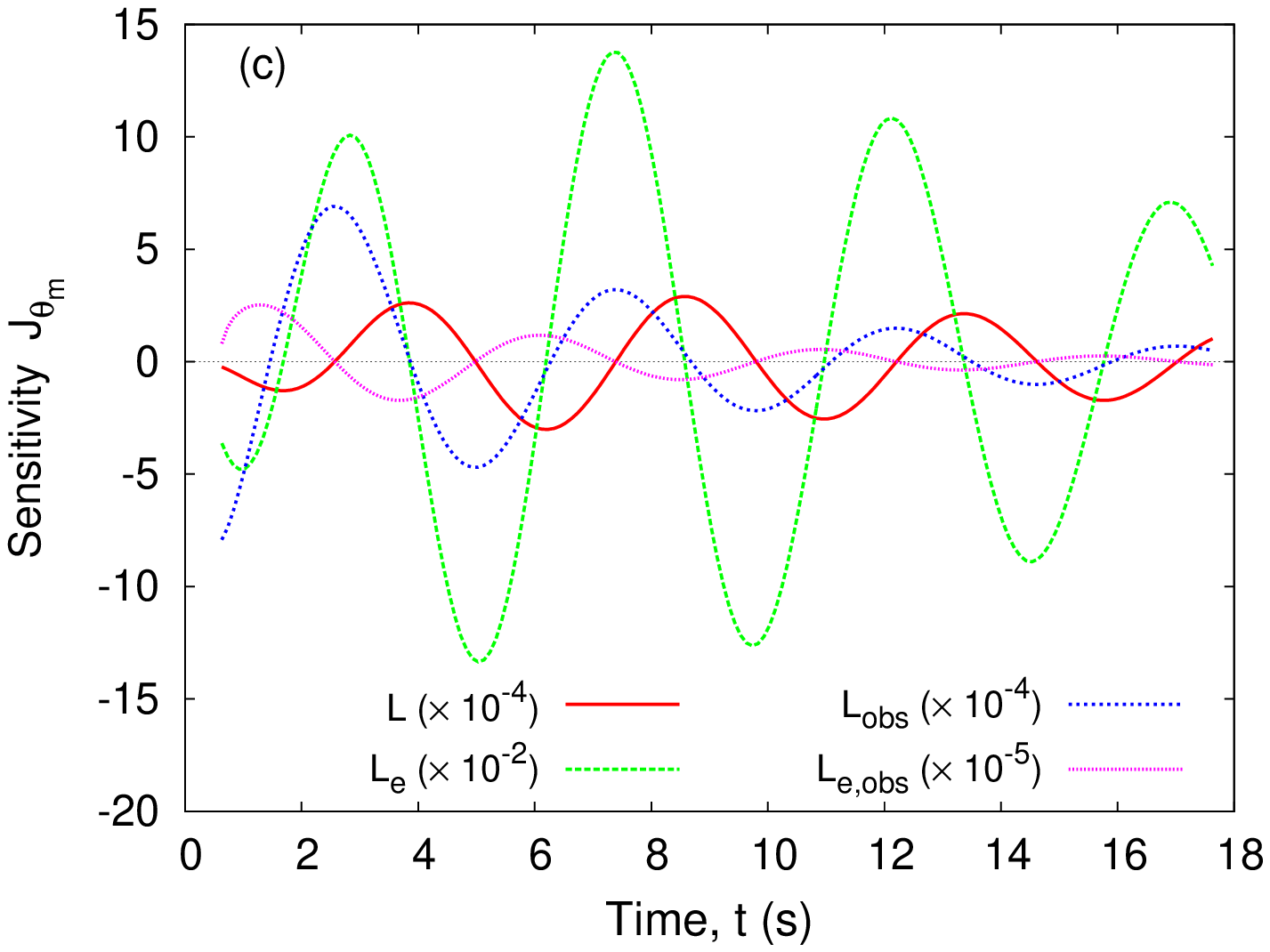}
\includegraphics[width=0.48\textwidth]{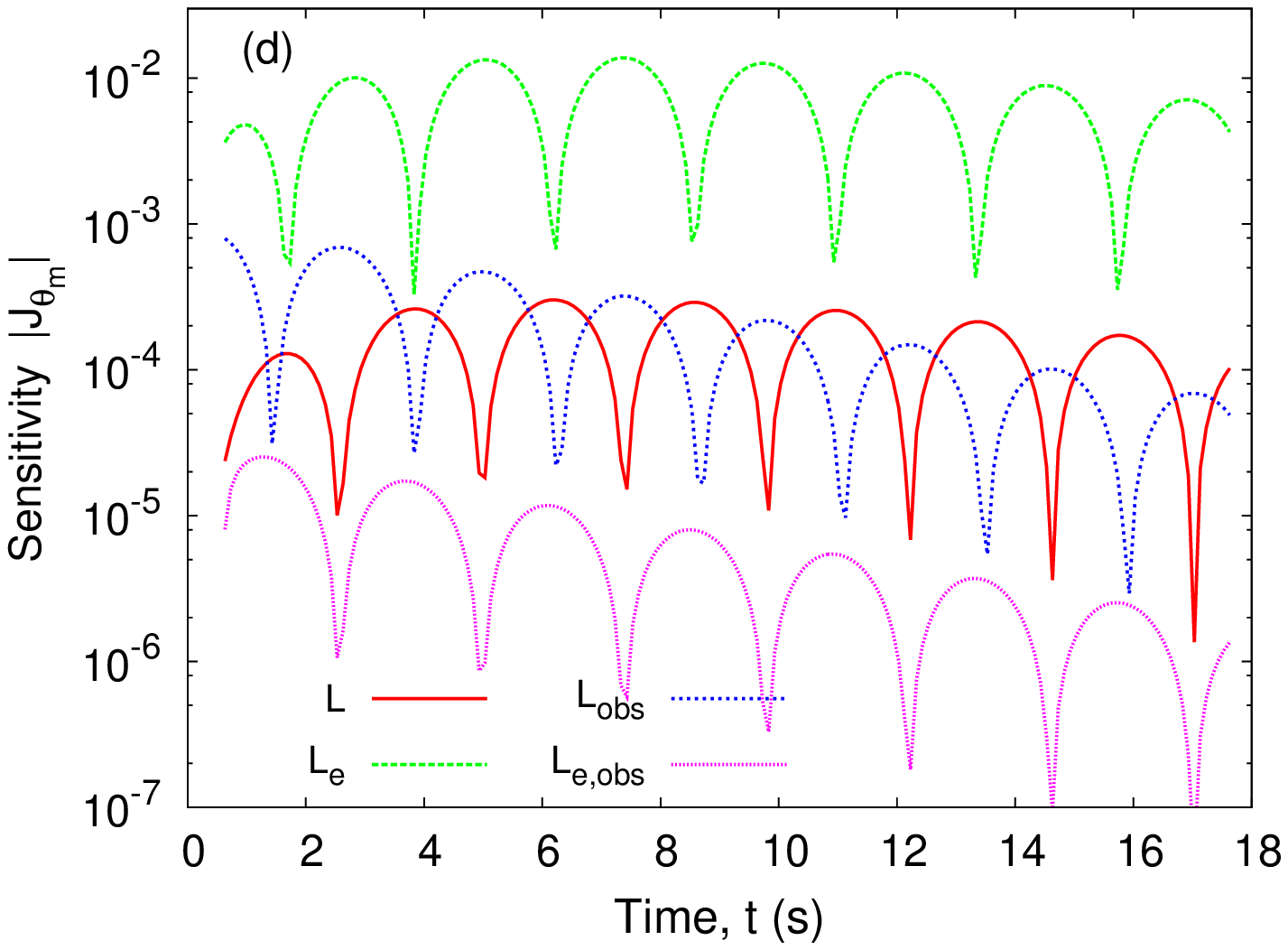}
\caption{\label{fig:jac1-1} Temporal variation of the sensitivity coefficients (linear scale (a \& c) and log scale (b \& d)) for the indicated parameters at the source-observation pair P13-MC1-1 ($5.1$~m below watertable). Subplot (a) shows observed response.}
\end{figure}

\begin{figure}[h] % Figure 6
\includegraphics[width=0.48\textwidth]{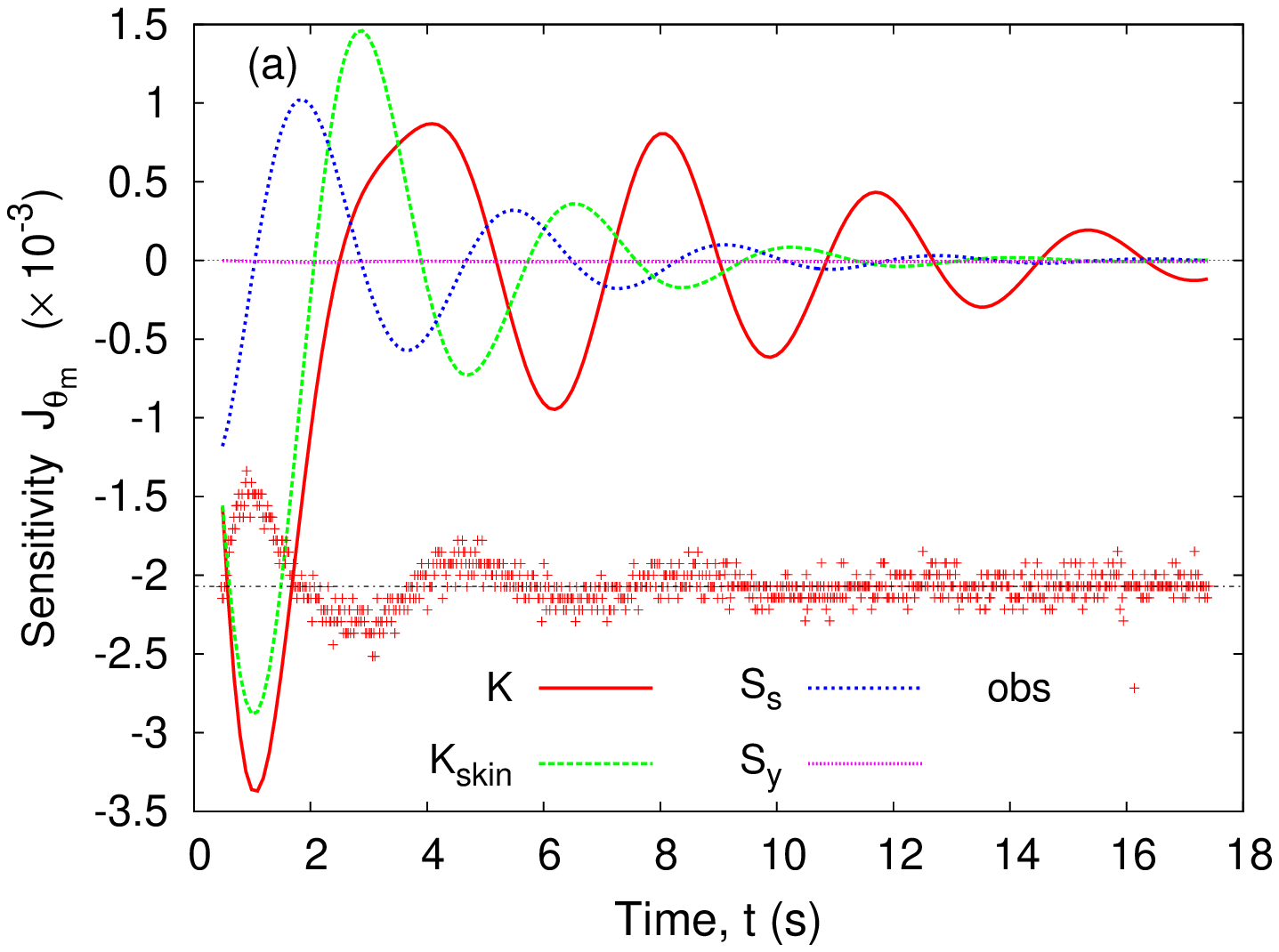}
\includegraphics[width=0.48\textwidth]{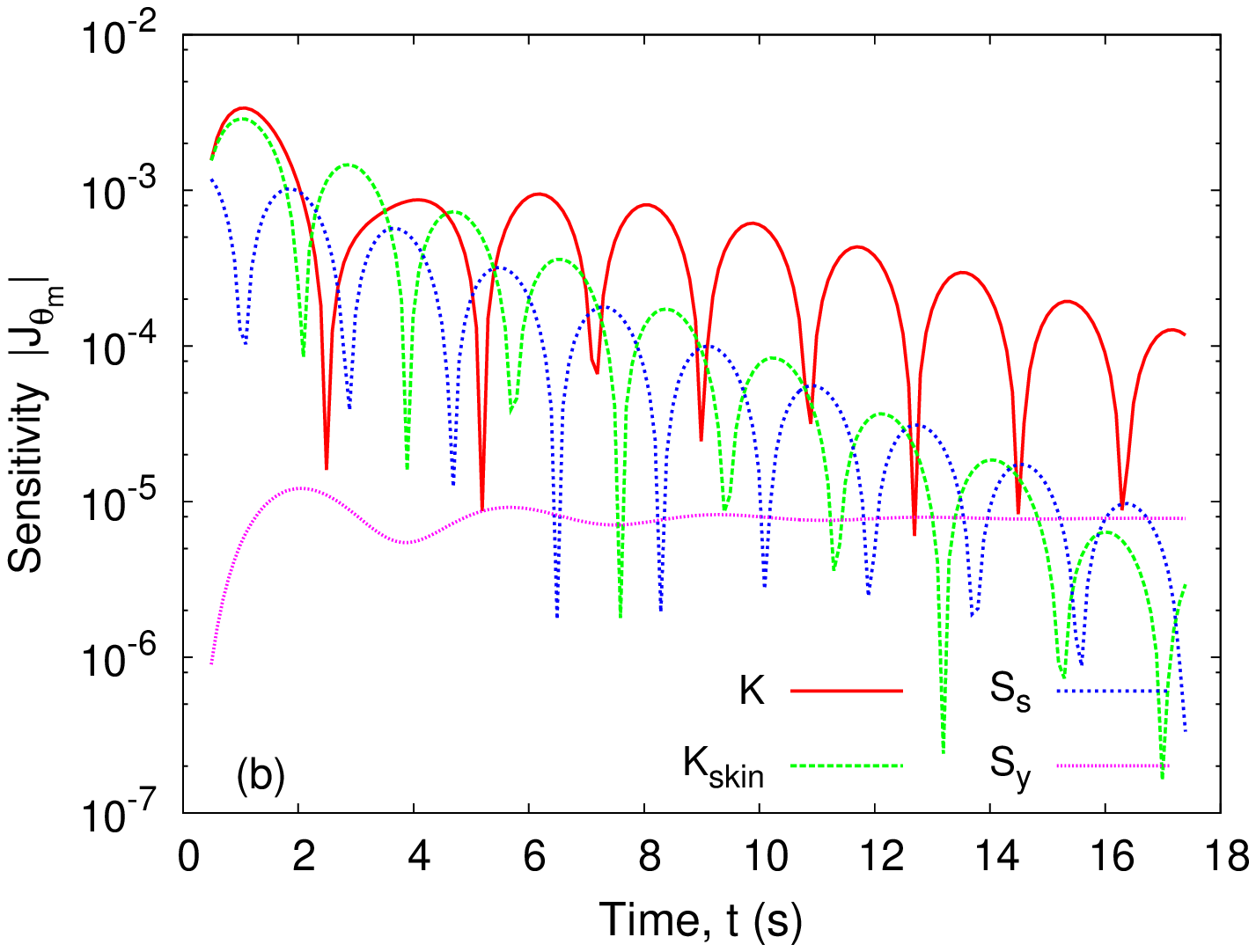}
\includegraphics[width=0.48\textwidth]{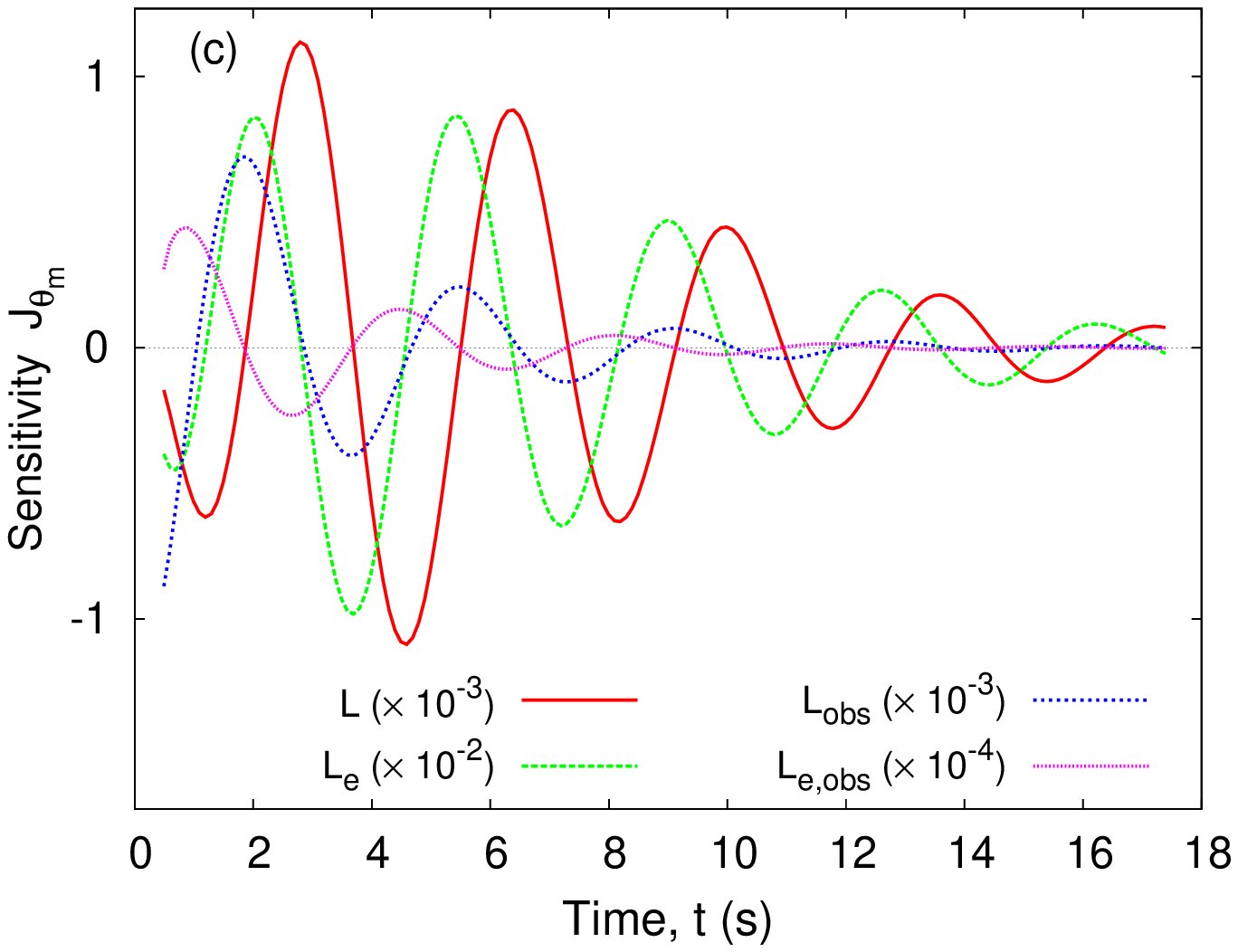}
\includegraphics[width=0.48\textwidth]{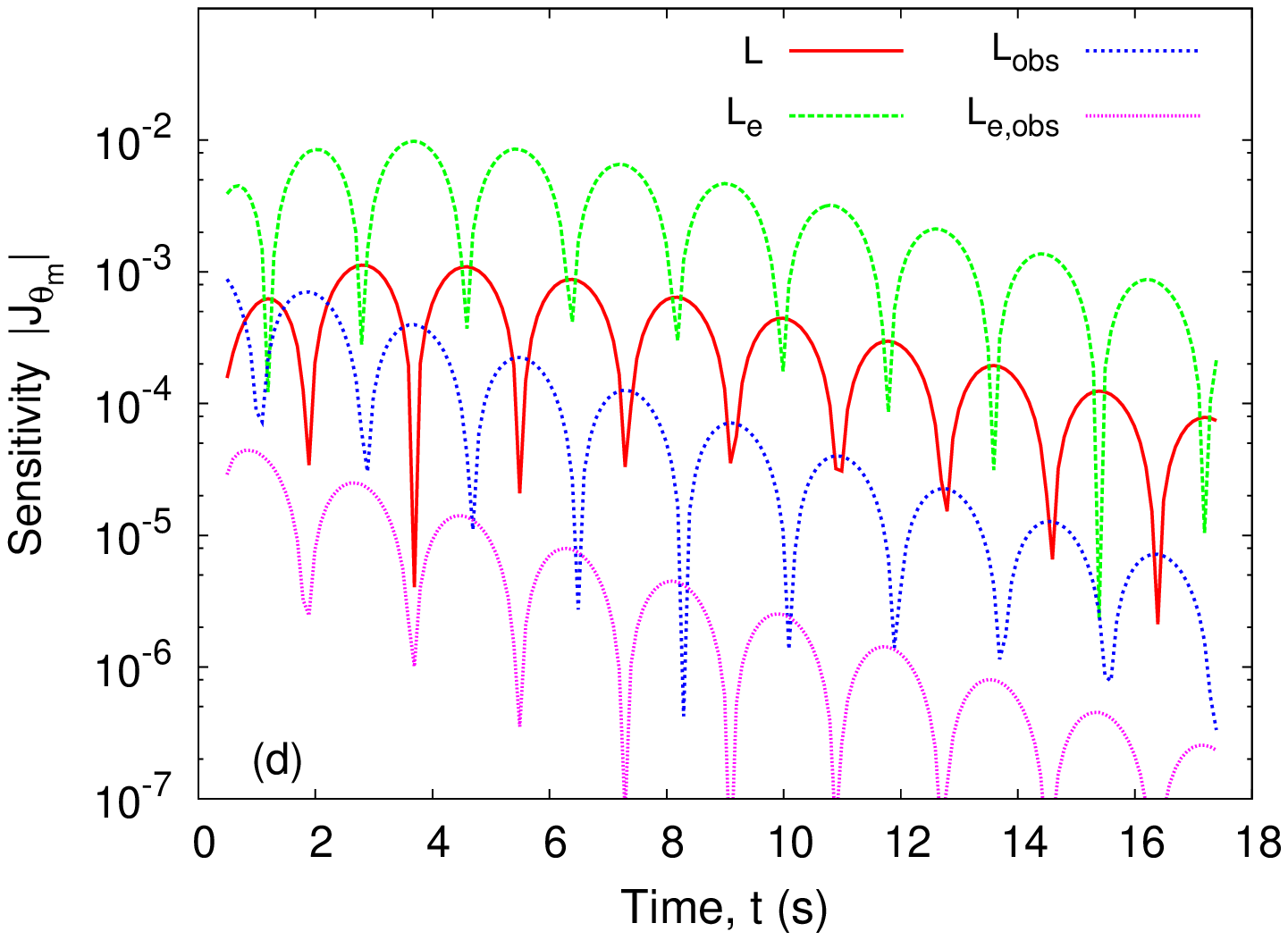}
\caption{\label{fig:jac1-9} Temporal variation of the sensitivity coefficients (linear scale (a \& c) and log scale (b \& d)) for the indicated parameters at the source-observation pair P13-MC1-5 ($3.1$~m below watertable). Subplot (a) shows observed response (same scale as response in Figure~\ref{fig:jac1-1}).}
\end{figure}

The model sensitivity or Jacobian matrix, $\mathbf{J}$, of dimensions $N\times M$, where $N$ is the number of observations and $M$ is the number of estimated parameters, is of central importance to parameter estimation. The sensitivity coefficients are simply the elements of the Jacobian matrix; they are the partial derivatives of the model-predicted aquifer head response, $s$, with respect to the estimated parameter $\theta_m$. Sensitivity coefficients are represented here as functions of time using the nomenclature 
\begin{equation}
J_{\theta_m}(t) = \frac{\partial s_{D,\mathrm{obs}}}{\partial \theta_m},
\end{equation}
where $m=1,2,...,M$. They describe the sensitivity of predicted model behavior (head response) to the model parameters. They provide a measure of the ease of estimation (identifiability) of the parameters from system state observations \citep{jacquez85}. The Jacobian matrix $\mathbf{J}$ has to satisfy the identifiability condition, $|\mathbf{J}^T\mathbf{J}| \neq 0$, for parameters to be estimable. This condition is typically satisfied for linearly independent sensitivity coefficients with appreciably large magnitudes. For this work, the number of parameters estimated is $M=8$, and the vector of estimated parameters is
\begin{equation}
\left(\theta_1,...,\theta_8 \right) = \left(K, K_\mathrm{skin}, S_s, S_y, L, L_e, L_\mathrm{obs}, L_{e,\mathrm{obs}} \right).
\end{equation}
Sensitivity coefficients for tests P13-MC1-1 (deepest) and P13-MC1-5 (intermediate depth) are shown as functions of time in Figures~\ref{fig:jac1-1} and \ref{fig:jac1-9}. Semi-log plots of the same information are included to more clearly show the non-zero sensitivity values at late-time. 

Generally, the sensitivities are oscillatory functions of time with decaying amplitudes that vary over several orders of magnitude among the parameters. Figure \ref{fig:jac1-1}(a) shows the sensitivity to the parameters $K$, $K_\mathrm{skin}$, $S_s$, and $S_y$. It is clear that well skin conductivity, $K_\mathrm{skin}$, has the highest peak sensitivity at early-time, and is therefore the most easily identifiable parameter from early-time data. Specific yield, $S_y$, has the smallest sensitivities (about an order of magnitude smaller than $K_\mathrm{skin}$) and was the least identifiable (most difficult to estimate) of all the parameters.

Figures~\ref{fig:jac1-1}(a) and (b) also show that the sensitivity functions are generally out of phase with each other as well as with the observed response. For example, the sensitivity function $J_{K}(t)$ is almost completely out of phase (phase-shift of $\sim\pi$) with $J_{K_\mathrm{skin}}$. The same is true for $J_{S_s}(t)$ and $J_{S_y}(t)$. This indicates linear-independence of the sensitivity coefficient among all four parameters. This is desirable as it implies that the identifiability condition is satisfied, permitting concomitant estimation of all these four parameters.

Figure~\ref{fig:jac1-1}(a) shows the $J_{S_y}$ is oscillatory with the small amplitudes and does not change sign, but decay more slowly than the other sensitivity responses. The predicted model response showed only modest sensitivity to specific yield, $S_y$, but the sensitivity becomes appreciably dominant at late-time (Figure~\ref{fig:jac1-1}(b)). \citet{malama2011} showed that slug tests are more sensitive to $S_y$ at late-time, and for relatively large initial perturbation. At late-time slug test head data are typically of low SNR (diminished data quality) making it difficult to discern effects of specific yield. However, with measurements such as those reported in \citet{malama2011} for a site in Montana, it is possible to obtain single-well slug tests data with clear effects due to $S_y$. The cross-hole slug test data analysed herein showed only modest watertable effects and the late-time data were not of sufficient quality. This suggests the importance of late-time data to maximize $S_y$ identifiability and estimability as also noted in \citet{malama2011}.

Figures~\ref{fig:jac1-1}(c) and (d) show scaled slug response sensitivities to parameters $L$, $L_e$, $L_\mathrm{obs}$, and $L_{e,\mathrm{obs}}$. They show orders of magnitude of variability, with sensitivity $L_e$ being three orders of magnitude larger than sensitivity to $L_{e,\mathrm{obs}}$. Whereas those of the parameters $L$, $L_e$, and $L_{e,\mathrm{obs}}$ are linearly independent (not of the same phase), the pair $L_e$ and $L_\mathrm{obs}$ are only linearly independent at very early time; they oscillate with the same phase after about 4 seconds. This illustrates a long temporal record of observations would not improve the joint estimation of these two parameters.

Figure~\ref{fig:jac1-9} shows the same information as depicted in Figure~\ref{fig:jac1-1} for a more damped observation location closer to the watertable. Model sensitivity to $K$ is equal to or larger than $K_\mathrm{skin}$ for this interval. Sensitivity to $S_s$ is also higher at early-time. Among parameters $K$, $K_\mathrm{skin}$, $S_s$, and $S_y$, sensitivity to $S_y$ is the smallest at early time (Figure~\ref{fig:jac1-9}(b)). The sensitivity to $S_y$ stays approximately constant with time after the first 10 second of the test, while sensitivities to $K$, $K_\mathrm{skin}$, and $S_s$ continue to decrease. It should be noted however, that the unfavorable SNR (low data quality) makes it very difficult to estimate $S_y$ from late-time data. Collecting data at 3.11~m below the watertable did not yield an appreciable improvement in specific yield identifiability over the interval at 5.11~m depth in Figure~\ref{fig:jac1-1}. The behavior depicted in Figure \ref{fig:jac1-9} also suggests only data collected in the first 12 seconds of the test are needed to estimate model parameters at this depth. The sensitivity coefficients for all but $K$ essentially vanish after about 12 seconds and the identifiability condition is no longer satisfied. Additionally, even for the case where the sensitivity coefficients appear to be in phase (linearly dependent) at early-time (compare $J_{K}(t)$ and $J_{K_\mathrm{skin}}(t)$ for $t\leq 2$ s for test P13-MC1-5), they quickly (in the first 12 s) become linearly independent with time. This again indicates that a temporal record of the response longer than a few seconds is sufficient for joint estimation of these two parameters.

\section{Conclusions}
Cross-hole slug test data were analysed with an extended version of the model of \citet{malama2011}. The semi-analytical model was modified for:
\begin{enumerate}
\item predicting heads at observation wells,
\item inclusion of borehole skin effects, 
\item use of the finite Hankel transform for computation expediency, and
\item inclusion of observation well storage and inertial effects.
\end{enumerate}
Estimates were obtained of formation and source/observation well skin hydraulic conductivity, specific storage, specific yield, and well characteristics that control oscillation frequency and degree of damping. The aim of the study was to evaluate the use of cross-hole slug test data to characterize vertical unconfined aquifer heterogeneity and understand identifiability and estimability of these parameters, especially specific yield. Estimated values of hydraulic conductivity and specific storage from PEST are indicative of a heterogeneous sand and gravel aquifer. Parameter estimation and sensitivity analysis show the model has effectively linearly independent sensitivity coefficients with respect to seven of the eight parameters estimated. These parameters are clearly jointly estimable from the data over the duration of the tests. It should be understood that the model used in this analysis was developed for a homogeneous but anisotropic aquifer and is thus of only limited and approximate applicability to analysis of a heterogeneous system.

Of the parameters estimated, model predictions were least sensitive to specific yield even near the watertable, which implies it was the least identifiable parameter. This is due to a combination of factors, including 
\begin{enumerate}
\item the short duration of the data record due to rapid signal decay with time ($<20$ seconds);
\item the increasing damping observed in monitoring locations near the watertable (resulting in even shorter temporal records), and;
\item the decreasing signal strength near the watertable, resulting in a lower signal-to-noise.
\end{enumerate}
The sensitivity function with respect to specific yield shows a relatively modest increase in magnitude with time (model sensitivity to the other model parameters tends to decrease, while that of $S_y$  asymptotically tends to a non-zero constant value), suggesting the importance of late-time data to improve its estimation. The analysis of \cite{malama2011} also indicated that the largest effect of specific yield on slug test response is at late-time, at which time the amplitude of the signal has decayed significantly in magnitude and quality. The absence of good quality late-time observations and the relative low sensitivities of specific yield explain the the wide variability of the estimates of $S_y$.

An important shortcoming of using cross-hole slug tests to characterize heterogeneity, as has been suggested in several field \citep{brauchler2010, brauchler2011} and synthetic \citep{paradis2015} hydraulic tomography studies, is the significant decay of the signal with distance from the source well and close to the water-table. These lead to low quality observations with low signal-to-noise ratios (SNR), and would require test redesign to improve parameter identifiability and estimability. One approach to change test design is to conduct tests with a sufficiently large initial displacement in the source well to achieve favorable SNR at late-time in the observation wells. This may, however, introduce non-linearities and potentially increase the importance of unsaturated flow above the watertable \citep{mishra2011saturated}. Another approach is to use more sensitive and low-noise pressure sensors, which would increase costs significantly, especially in the cross-hole multilevel testing set-up where a large network of sensors is used for data acquisition. This would be particularly useful close to the watertable and further from the source well due to significant signal strength decay decline. This decline in signal strength limits the usefulness of crosshole slug tests for large-scale aquifer characterization using hydraulic tomography. Lastly, conducting multiple test repetitions and stacking the response data, akin to seismic data stacking \citep{jones1987}, can be used to amplify signal and increase the SNR.

\section*{Acknowledgements}
Sandia National Laboratories is a multi-program laboratory managed and operated by Sandia Corporation, a wholly owned subsidiary of Lockheed Martin Corporation, for the U.S. Department of Energy's National Nuclear Security Administration under contract DE-AC04-94AL85000.

\renewcommand{\theequation}{A-\arabic{equation}}
\setcounter{equation}{0}
\section*{Appendix A: Solution with Linearized Watertable Kinematic Condition}
\label{sec:appendixa}
The solution can be written in dimensionless form  for the intervals above, below, and across from the source well completion interval as
\begin{equation}
\label{eqn:}
s_D = \left\{
  \begin{split}
   s_D^{(1)} & \quad \mbox{$\forall z_D\in[0,d_D]$}\\
   s_D^{(2)} & \quad \mbox{$\forall z_D\in[d_D,l_D]$}\\
   s_D^{(3)} & \quad \mbox{$\forall z_D\in[l_D,1]$},
  \end{split} \right.
\end{equation}
where $s_D^{(n)}$ solves
\begin{equation}
\label{eqn:unconfinedPDE}
\frac{\partial s_D^{(n)}}{\partial t_D} = \frac{1}{r}\frac{\partial}{\partial r_D} \left(r_D \frac{\partial s_D^{(n)}}{\partial r_D}\right) + \kappa \frac{\partial^2 s_D^{(n)}}{\partial z_D^2}.
\end{equation}
The initial and boundary conditions are
\begin{equation}
\label{eqn:initialfarfieldBC2}
s_D^{(n)} (t_D=0) = s_D^{(n)} (r_D=R_D) = 0
\end{equation}
\begin{equation}
\label{eqn:centerBC}
\lim_{r_D\rightarrow 0} r_D\frac{\partial s_D^{(1)}}{\partial r_D} = \lim_{r_D\rightarrow 0} r_D \frac{\partial s_D^{(3)}}{\partial r_D} = 0
\end{equation}
\begin{equation}
\label{eqn:kinematicBCD}
\left.\frac{\partial s_D^{(1)}}{\partial z_D} \right|_{z_D=0} = \frac{1}{\alpha_D}\left.\frac{\partial s_D^{(1)}}{\partial t_D} \right|_{z_D=0}
\end{equation}
\begin{equation}
\label{eqn:noflow}
\left.\frac{\partial s_D^{(3)}}{\partial z_D} \right|_{z_D=1} = 0
\end{equation}
\begin{equation}
\label{eqn:massD}
\left. r_D\frac{\partial s_D^{(2)}}{\partial r_D}\right|_{r_D = r_{D,w}} = C_D \frac{\mathrm{d} \Phi_\mathrm{uc}}{\mathrm{d}t_D},
\end{equation}
\begin{equation}
\Phi_\mathrm{uc}(t_D = 0) = 1,
\end{equation}
and
\begin{equation}
\label{eqn:momentumDAp}
\beta_2 \frac{\mathrm{d}^2\Phi_\mathrm{uc}}{\mathrm{d}t_D^2} + \beta_1 \frac{\mathrm{d}\Phi_\mathrm{uc}}{\mathrm{d}t_D} +\Phi_\mathrm{uc}=\frac{1}{b_D}\int_{d_D}^{l_D} s_D^{(2)}(r_{D,w},z_D,t_D) \;\mathrm{d}z_D.
\end{equation}
Additionally, continuity of head and flux is imposed at $z_D=d_D$ and $z_D=l_D$ via
\begin{equation}
\label{eqn:s_continuity1}
s_D^{(1)} (t_D,r_D,z_D=d_D) = s_D^{(2)} (z_D=d_D),
\end{equation}
\begin{equation}
\label{eqn:ds_continuity1}
\left. \frac{\partial s_D^{(1)}}{\partial z_D} \right|_{z_D=d_D} = \left. \frac{\partial s_D^{(2)}}{\partial z_D} \right|_{z_D=d_D},
\end{equation}
\begin{equation}
s_D^{(3)} (t_D,r_D,z_D=l_D) = s_D^{(2)} (z_D=l_D),
\end{equation}
and
\begin{equation}
\left. \frac{\partial s_D^{(3)}}{\partial z_D} \right|_{z_D=l_D} = \left. \frac{\partial s_D^{(2)}}{\partial z_D} \right|_{z_D=l_D}.
\end{equation}
This flow problem is solved using Laplace and Hankel transforms. Taking the Laplace and Hankel transforms of Equation (\ref{eqn:unconfinedPDE}) for $n=1,3$, and taking into account the initial and boundary conditions in (\ref{eqn:initialfarfieldBC2}) and (\ref{eqn:centerBC}), gives the ordinary differential equation
\begin{equation}
\label{eqn:n1n3ODE}
\frac{\mathrm{d}^2 \hat{\overline{s}}_D^{(n)}}{\mathrm{d} z_D^2} - \eta^2 \hat{\overline{s}}_D^{(n)} = 0
\end{equation}
where $\hat{\overline{s}}_D^{(n)} = \mathsf{H}\{\mathcal{L}\{s_D^{(n)}\}\}$ is the double Laplace-Hankel transform of the function $s_D^{(n)}$, $\eta^2 = (p+a_i^2)/\kappa$, and $p$ and $a_i$ are the Laplace and finite Hankel transform parameters, respectively. Equation (\ref{eqn:n1n3ODE}) has the general solution
\begin{equation}
\hat{\overline{s}}_D^{(n)} = A_n e^{\eta z_D} + B_n e^{-\eta z_D}.
\end{equation}
The boundary condition at the watertable, Equation (\ref{eqn:kinematicBCD}), in Laplace--Hankel transform space, becomes
\begin{equation}
\label{eqn:kinematicBCD}
\left.\frac{\partial \hat{\overline{s}}_D^{(1)}}{\partial z_D} \right|_{z_D=0} = \frac{p}{\alpha_D} \hat{\overline{s}}_D^{(1)}(z_D=0).
\end{equation}
Applying this boundary condition leads to
\begin{equation}
\label{eqn:AB1}
(1-\varepsilon)A_1 - (1+\varepsilon)B_1 = 0,
\end{equation}
where $\varepsilon=p/(\eta\alpha_D)$. Applying the continuity conditions at $z_D=d_D$ (Equations (\ref{eqn:s_continuity1}) and (\ref{eqn:ds_continuity1})), lead to
\begin{equation}
\label{eqn:AB2}
A_1 e^{\eta d_D} + B_1 e^{-\eta d_D} = \hat{\overline{s}}_D^{(2)} (z_D=d_D),
\end{equation}
and
\begin{equation}
\label{eqn:AB3}
\eta\left(A_1 e^{\eta d_D} - B_1 e^{-\eta d_D}\right) = \left.\frac{\mathrm{d} \hat{\overline{s}}_D^{(2)}}{\mathrm{d} z_D} \right|_{z_D=d_D}.
\end{equation}
Similarly, applying the no flow boundary condition at $z_D=1$ (Equation \ref{eqn:noflow}), leads to
\begin{equation}
\hat{\overline{s}}_D^{(3)} = 2B_3 e^{-\eta} \cosh\left(\eta z_D^\ast\right),
\end{equation}
where $z_D^\ast = 1-z_D$. Continuity conditions at $z_D=l_D$ lead to
\begin{equation}
\label{eqn:AB4}
2B_3 e^{-\eta} \cosh\left (\eta l_D^\ast \right) = \hat{\overline{s}}_D^{(2)} (z_D=l_D),
\end{equation}
\begin{equation}
\label{eqn:AB5}
-2\eta B_3 e^{-\eta} \sinh\left(\eta l_D^\ast\right) = \left.\frac{\mathrm{d} \hat{\overline{s}}_D^{(2)}}{\mathrm{d} z_D} \right|_{z_D=l_D},
\end{equation}
where  $l_D^\ast = 1 - l_D$ and $d_D^\ast = 1-d_D$.
For $n=2$, solving Equation (\ref{eqn:unconfinedPDE}) in Laplace-Hankel transform space yields
\begin{equation}
\label{eqn:sD2solution}
\hat{\overline{s}}_D^{(2)} = \hat{\overline{u}}_D + \hat{\overline{v}}_D,
\end{equation}
where
\begin{equation}
\hat{\overline{u}}_D = \frac{C_D(1-p\overline{\Phi}_\mathrm{uc})}{\kappa\eta^2\xi_w\mathrm{K}_1(\xi_w)},
\end{equation}
and
\begin{equation}
\label{eqn:vDsolution}
 \hat{\overline{v}}_D = A_2 e^{\eta z_D} + B_2 e^{-\eta z_D}.
\end{equation}
The five equations (\ref{eqn:AB1})--(\ref{eqn:AB3}), (\ref{eqn:AB4}) and (\ref{eqn:AB5}), together with Equation (\ref{eqn:sD2solution}) can be used to determine the five unknown coefficients $A_1$, $A_2$, and $B_1$--$B_3$. It can then be shown that
\begin{equation}
\label{eqn:vDfunction}
 \hat{\overline{v}}_D = -\frac{\hat{\overline{u}}_D}{\Delta_0} \left\{\Delta_1 \cosh\left(\eta z_D^\ast\right) + \sinh\left(\eta \l_D^\ast\right) \left[\cosh\left(\eta z_D\right) + \varepsilon \sinh\left(\eta z_D\right)\right] \right\}.
\end{equation}
The integral in Equation (\ref{eqn:momentumDAp}) is
\begin{equation}
\label{eqn:average_sD2}
\begin{split}
\frac{1}{b_D}\int_{d_D}^{l_D} \hat{\overline{s}}_D^{(2)} \;\mathrm{d}z_D & = \hat{\overline{u}}_D + \frac{1}{b_D} \int_{d_D}^{l_D} \hat{\overline{v}}_D \;\mathrm{d}z_D\\
& = \hat{\overline{u}}_D + \left\langle \hat{\overline{v}}_D \right\rangle.
\end{split}
\end{equation}
Substituting Equation (\ref{eqn:vDfunction}) into (\ref{eqn:average_sD2}) leads to 
\begin{equation}
\label{eqn:average_sD2b}
\frac{1}{b_D}\int_{d_D}^{l_D} \hat{\overline{s}}_D^{(2)} \mathrm{d}z_D  =\hat{\overline{u}}_D\left(1-\left\langle \hat{\overline{w}}_D \right\rangle\right),
\end{equation}
where
\begin{equation}
\label{eqn:wD}
\begin{split}
\langle \hat{\overline{w}}_D \rangle &= \frac{1}{b_D\eta \Delta_0} \left[\Delta_1 \sinh\left(\eta d_D^\ast\right) + \left(\Delta_2 - 2\Delta_1\right) \sinh\left(\eta l_D^\ast\right)\right],\\
\Delta_0 &= \sinh(\eta) + \varepsilon \cosh(\eta),\\
\Delta_1 &= \sinh(\eta d_D) + \varepsilon \cosh(\eta d_D),\\
\Delta_2 &= \sinh(\eta l_D) + \varepsilon \cosh(\eta l_D).
\end{split}
\end{equation}

Taking the Laplace transform of (\ref{eqn:momentumDAp}) and replacing the integral on the left-hand-side with (\ref{eqn:average_sD2b}), gives
\begin{equation}
\label{eq:phiuc2}
(p^2 + \beta_1p +\beta_2)\overline{\Phi}_\mathrm{uc} - p - \beta_1 = \frac{1}{2}\left(1-p\overline{\Phi}_\mathrm{uc} \right) \overline{\Omega}
\end{equation}
where $\hat{\overline{\Omega}}$ is defined in (\ref{eqn:HankLap_Omega}). Solving \eqref{eq:phiuc2} for $\overline{\Phi}_\mathrm{uc}$ yields the required source well response in Laplace space.

\section*{Notation}
\begin{tabular}{llc}
$a_i$ & finite Hankel transform parameter & $\mathrm{-}$\\
$B$ & Aquifer initial thickness & $\mathrm{L}$ \\ 
$b_s$ & Length of source well test interval & $\mathrm{L}$ \\ 
$C_w$ & Coefficient of wellbore storage &  $\mathrm{L^2}$ \\ 
$d/d_\mathrm{o}$ & Depth of top of source/observation well test interval below watertable & $\mathrm{L}$ \\ 
$g$ & Acceleration due to gravity & $\mathrm{L \cdot T^{-2}}$ \\ 
$H$ & Hydraulic head change from equilibrium position in source well & $\mathrm{L}$ \\
$H_0$ & Initial slug input & $\mathrm{L}$ \\ 
$K$ & Formation hydraulic conductivity & $\mathrm{L \cdot T^{-1}}$ \\ 
$K_r$ & Radial formation hydraulic conductivity & $\mathrm{L \cdot T^{-1}}$ \\ 
$K_{z}$ & Vertical formation hydraulic conductivity & $\mathrm{L \cdot T^{-1}}$ \\ 
$K_\mathrm{skin}$ & Skin hydraulic conductivity & $\mathrm{L \cdot T^{-1}}$ \\ 
$l/l_\mathrm{o}$ & Depth of bottom of source/observation well test interval below watertable & $\mathrm{L}$ \\ 
$L/L_\mathrm{obs}$ & Characteristic length for source/observation well damping term & $\mathrm{L}$ \\ 
$L_e/L_{e,\mathrm{obs}}$ & Characteristic length for source/observation well oscillatory term & $\mathrm{L}$ \\ 
$p$ & Laplace transform parameter  & $\mathrm{-}$\\
$r$ & Radial coordinate, out from center of source well & $\mathrm{L}$ \\ 
$R$ & Domain radius, out from center of source well & $\mathrm{L}$ \\ 
$r_c$ & Radius of source well tubing at water-table & $\mathrm{L}$ \\ 
$r_w$ & Radius of source well at test interval & $\mathrm{L}$ \\ 
$s$ & Hydraulic head change from initial conditions & $\mathrm{L}$ \\
$S_{s}$ & Specific storage & $\mathrm{L^{-1}}$\\ 
$S_y$ & Specific yield & $-$ \\ 
$t$ & Time since slug initiation & $\mathrm{T}$ \\
$T_c$ & Characteristic time ($T_c=B^2/\alpha_{r,1}$) & $\mathrm{T}$ \\
$z$ & Vertical coordinate, down from water-table & $\mathrm{L}$  \\
$\alpha_{r,i}$ & Hydraulic diffusivity of $i^\mathrm{th}$ zone & $\mathrm{L^2 \cdot T^{-1}}$ \\ 
$\gamma_s$ & Source well damping coefficient & $\mathrm{T^{-1}}$ \\ 
$\nu$ & Kinematic viscosity of water & $\mathrm{L^2 \cdot T^{-1}}$ \\ 
\end{tabular} 

\section*{References}
%%\bibliographystyle{elsarticle-harv} 
%%\bibliography{hydrology}

%%\end{linenumbers}

\end{document}